\RequirePackage{fix-cm}
\documentclass{svjour3}                     
\smartqed  
\usepackage{subfigure}
\usepackage{xspace}
\usepackage{amssymb, amsmath}
\usepackage{color}
\usepackage{colortbl}
\usepackage{graphicx}
\usepackage[linesnumbered,ruled]{algorithm2e}
\usepackage{lineno}

\usepackage{dirtytalk}
\usepackage[utf8]{inputenc}
\usepackage[T1]{fontenc}

\newcommand{\thiago}[1]{{\color{black}#1}}

\newcommand{\willi}[1]{{\color{black}#1}}

\newcommand{\pe}[1]{{\color{black}#1}}

\newcommand{\novo}[1]{{\color{black}#1}}

\begin{document}

\title{Gender Matters! Analyzing Global Cultural Gender Preferences for Venues Using Social Sensing}

\titlerunning{Gender Matters!}        

\author{*Willi Mueller\thanks{*Worked equally in this study.} \and
        *Thiago H Silva \and
        Jussara M Almeida \and
        Antonio A F Loureiro
}


\institute{T. H. Silva, J. M. Almeida, and A. A. F. Loureiro \at
              Department of Computer Science, Universidade Federal de Minas Gerais, Belo Horizonte, MG, Brazil \\
              Tel.: +55-31-3409-5860\\
              Fax: +55-31-3409-5858\\
              \email{thiagohs,jussara,loureiro@dcc.ufmg.br} \\
             \emph{Present address:} of T. Silva is at Department of Informatics, Universidade Tecnol\'{o}gica Federal do Paran\'{a}, Curitiba, PR, Brazil\\
           \and
           W. Mueller \at
             Department of Computer Science, Hasso-Plattner-Institut, Potsdam, Brandenburg, Germany\\
             \email{willi.mueller@student.hpi.de}
}


\maketitle

\begin{abstract}

Gender differences  is a phenomenon around the world actively researched by social scientists. Traditionally,  the data used to support such studies is manually obtained, often through surveys with volunteers. However, due to their inherent high costs  because of manual steps, such traditional methods do not quickly scale to large-size  studies. We here investigate a particular aspect of gender differences: preferences for venues. To that end we explore the use of check-in data collected from Foursquare to estimate cultural gender preferences for venues in the physical world. For that, we first demonstrate that by analyzing the check-in data  in various regions of the world we can find significant differences in preferences for specific venues between gender groups. Some of these significant differences reflect well-known cultural patterns. Moreover, we also gathered evidence that our methodology offers useful information about gender preference for venues in a given region in the real world. This suggests that gender and venue preferences observed may not be independent. Our results suggests that our proposed methodology could be a promising tool to support studies on gender preferences for venues at different spatial granularities around the world, being faster and cheaper than traditional methods, besides quickly capturing changes in the real world.


\end{abstract}

\section{Introduction}

\thiago{Gender differences can be considered one of the great puzzles of modern society. It has a subjective nature, and may vary greatly across cultures~\cite{sen2001many,szymanowicz2011intelligent,harrison2000culture}. For instance, when comparing different regions of the world, women and men often differ in their assumed capacities, and others. This makes gender differences hard to explain. Indeed, over the past decades, this topic has received a lot of attention in the are of Social Science, but there is still a long way  to a consensus on the subject~\cite{hyde2005gender,ridgeway2011framed}.}

In order to study the differences between gender groups around the world,   social scientists often rely   on manual methods to gather heterogeneous data, often using surveys with volunteers. The collected data may  then be aggregated to compute particular metrics, such as the Gender Inequality Index (GII) developed by the United Nations Development Programme (UNDP)~\cite{undp2014}.

However, these traditional methods are time-consuming because of the manual steps. Moreover, data produced under such conditions are commonly released after long time intervals (e.g., it could take several years).  Therefore, they cannot   quickly capture changes in the dynamics of societies. Besides, the results from cross-regional gender differences studies, such as the GII reports, are usually available  only for large  geographic regions, often  countries. Thus, even though survey-based studies could be carried out in arbitrary small regions, such as a city, a neighborhood or even a particular venue (e.g., a university or a mall), information about gender differences at such fine spatial granularities is not easily available.

\thiago{With that, one of the main research questions of this paper is: Can we propose a complementary method to help in the study of gender differences in a large scale and in a faster way than traditional methods?}

Location-based social networks (LBSNs), such as Foursquare\footnote{http://www.foursquare.com.}, are currently very popular, mostly due to the widespread use of smartphones around the world. In such applications, users implicitly express their preferences for locations by performing check-ins at specific venues. Check-ins can then be seen as a source of \textit{social sensing}, capturing how people behave in the real world with respect to the places they often visit. As discussed in~\cite{silvaICWSM14,cranshaw:livehoods}, such signals can be explored to better understand human  dynamics in urban areas, and, particularly, culture-related urban patterns.

We focus on a particular aspect of the culture of a society, namely gender bias~\cite{baron2010talking,davidGarcia14,szymanowicz2011intelligent,harrison2000culture,vzivzek1989sublime,Magno2014,wagner2015s,Volkovich2014}. We aim at investigating whether user check-ins in LBSNs can also be used to assess \textit{cultural gender preferences for venues} at different urban regions of the physical world. \thiago{In our context, culture is expressed through preference for a particular venue.} To capture that, we propose a methodology to quantify the differences  between male and female users in preferences for particular venues. The aggregation of such differences over multiple venues could then be used, for example, in the construction of an indicator of gender differences in a given region.

We illustrate the use of our methodology by  extracting user preferences  for venues located in different urban  regions around the world from check-in data collected from Foursquare.  We then identify significant differences for specific venues between gender groups in various regions, which suggest that gender and venue preferences may not be independent in those regions.  We illustrate the potential use of our methodology by applying it to various spatial granularities, including countries, cities, and a particular type of venues in a given city.

We demonstrate one application that aims at identifying groups of similar urban areas according to the degree of gender preference for venues observed in different (types of) places located in those areas. \thiago{Furthermore, we investigate to which extent gender preferences for venues is related to gender differences. For that, we} compared our results with those produced using the United Nations GII values. This analysis suggests that our approach might capture \thiago{some} essential aspects  of gender differences. Besides, it also motivates the study of new approaches to using social sensing jointly with other data in future developments of gender differences indices.

In summary, the main contributions of this work are: (i) a methodology to characterize gender preferences for venues in different regions at different spatial granularities, around the world, based on LBSNs and (ii) a study of  our methodology as a means to assess cultural gender preferences for venues showing its potential for different studies in several areas.

The results  that our methodology produces could be a promising tool to support large-scale gender preferences for venues studies that require less human effort and time, compared with traditional methods, and  can quickly react to changes in the real world because it relies on LBSNs data.  The obtained results could be used in several contexts. For instance, they might help policy makers to evaluate the effect of implemented policies regarding the minimization of gender differences in certain regions/venues of the city.  Similarly, they might help business owners and marketers to better understand their  consumers. For example, \willi{if a coffee shop has a very distinct pattern of consumer gender compared with other coffee shops in the same city, the owner could exploit this knowledge  to promote advertisement}. \willi{Our method} may also be used to  identify similarities and discrepancies regarding venue preferences of  gender groups across different regions. Finally,  the results might drive the design of more culturally-aware venue recommender systems, as men and women may have different preferences in regions with distinct cultures.

The rest of this article is organized as follows. Section~\ref{secRelatedwork} review the related work. Section~\ref{secDatasetUsed} introduces our dataset, while Section~\ref{secCharacterization} presents a study about  gender preferences for venues in urban regions of different sizes. Section~\ref{secApplications} presents some applications that could benefit from our work. Section~\ref{secValidation} compares our results with official indices of gender differences. Section \ref{secLimitations} discusses some of the known limitations of our study. Finally, Section~\ref{conclusions} presents the concluding remarks and future work.

\section{Related Work}\label{secRelatedwork}

The study of gender differences has been receiving a considering amount of attention in different areas. Some recent studies  include the investigation of gender differences in education~\cite{buchmann2008gender},  in relationships \cite{van2011rationalising,szymanowicz2011intelligent}, and with respect to the use of technology~\cite{hargittai2008digital}. In the latter, the authors analyzed how 270 adults used the Web, aiming at identifying differences in online activity. These prior studies, as most social science studies, relied on surveys with a reasonably small sample size. However, such manual approach imposes big challenges to studies with larger sample sizes (e.g., thousands or millions of users).

Recently, scientists are jointly applying techniques from  Computer Science and  Statistics to support sociological studies using large-scale datasets. For example, Kershaw et al.~\cite{Kershaw2014} looked into the use of social media to monitor the rate of alcohol consumption. Weber et al.~\cite{Weber2012} used web search query logs to analyze and visualize political issues. Some other topics of study include the understanding of city dynamics~\cite{zambaldi2014lightweight,Silva2014toit,cranshaw:livehoods}, event detection/study~\cite{georgiev2014call,Sakaki2010,alsaedi2014combined,becker2011beyond,Pan20133,gomide2010dengue}, cultural differences~\cite{garcia2013,silvaICWSM14,Hochman:2012,brunoGoncalves2013,davidGarcia14}, and gender inference~\cite{ciot2013EMNLP,Burger2011,liu2013s}.

On the particular topic of cross-gender differences, Ottoni et al.~\cite{ottoni2013ladies}
observed a great difference between female and male users with respect to their motivations for using Pinterest. Lou et al.~\cite{Lou2013} investigated how  gender swapping is revealed in massively multiplayer online games, observing that both male and female players achieve higher levels in the game faster with a male avatar than with a female avatar. De Las Casas et al.~\cite{deLasCasas2014}  characterized the use of Google+ by members who declared themselves as neither female nor male individuals, but as {\it other}. Cunha et al.~\cite{Cunha2012} studied gender distinctions in the usage of Twitter hashtags, concluding that  gender can be considered a social factor that influences the user's choice of particular hashtags about a given topic. Garcia et al.~\cite{davidGarcia14} measured gender biases of dialogues in movies and social media, showing that Twitter presents a male bias, whereas MySpace does not. Wagner et al. \cite{wagner2015s} present a method for assessing gender bias on Wikipedia. Gender bias in Wikipedia is also studied by Graells-Garrido et al \cite{Graells-Garrido:2015}. Magno and Weber \cite{Magno2014} study gender inequality through user participation in two online social networks, Twitter and Google+, finding, for example, that the gap between the number of users correlates with the gender gap index, i.e., countries with more men than women online are countries with higher gender difference. Volkovich et al. \cite{Volkovich2014} also study gender difference in a large online social network, looking mainly in the way how men and women sign up to a social network platform and make friends online. They found a general tendency towards gender homophily, more marked for women.

In this work, we also use a large-scale dataset, in our case data from a popular LBSN, which expresses user preferences for venues in a region, for various regions around the globe. However, unlike the aforementioned prior studies, we want to infer relevant cross-gender differences in the physical world, instead of online. To that end, we propose a methodology to quantify the differences between male and female users in preferences for particular venues across different cultures.

\section{Dataset Description}\label{secDatasetUsed}

A common approach to conducting studies on human behavior  is  by means of surveys, where participants answer questions administered through interviews or questionnaires~\cite{hofstede2001culture,jackson2011research,vianello1990gender}. However, despite its wide adoption, survey-based studies do have some severe constraints, which are well known to researchers. First, they may be costly and do not scale up. It is often hard to obtain data of millions or even thousands of people, particularly when focusing on multiple geographic regions. Second, they provide static information, reflecting human behavior at a specific point in time. Thus, they cannot capture well the natural changes we may expect from dynamic societies.

Instead of relying on survey data, we here investigate the use of publicly available data from LBSNs, notably Foursquare, to study gender preference for venues.  LBSNs can be accessed everywhere by anyone with an Internet connection, solving the scalability problem and allowing the collection of data from (potentially) the entire world \cite{Silva2014toit}. Moreover, these systems are quite dynamic, capturing behavioral changes of their users when they occur.

Nevertheless, the use of LBSN data also has some limitations, such as an inherent bias to regions and population groups where the application and required technology are more widely used. Yet, recent work has exploited  this type of data to support social studies on various topics, as further discussed in Section~\ref{secRelatedwork}. We here focus on gender, and investigate its use to drive studies on gender preferences for venues.

Specifically, our dataset consists of check-ins made by Foursquare users and become publicly available through Twitter between April $25^{th}$ and May $1^{st}$ 2014. This dataset contains roughly $2.9$ million tweets with check-ins shared by approximately $630$ thousands users. Foursquare venues are grouped into ten categories  (in parenthesis are the abbreviations used here): Arts \& Entertainment (Arts); College \& University (Education); Event; Food; Nightlife Spot (Nightlife); Outdoors \& Recreation; Professional \& Other Places (Work); Residence; Shop \& Service; Travel \& Transport. Each category, in turn, has several subcategories. \willi{For example, Comedy Club, Museum, and Casino are subcategories of Arts. Bar, Rock Club, and Pub are subcategories of Nightlife. College Lab, Fraternity House, and Student Center are subcategories of Education. Finally, Baseball Stadium, Surf Spot, and Park are subcategories of Outdoors \& Recreation.}

We applied the following filters to our dataset: We only considered check-ins performed by users who specified either ``male" or ``female" as   gender in their Foursquare profiles. We disregarded all check-ins in venues with fewer than five check-ins and considered only one check-in per user per venue to avoid users with many check-ins biasing the popularity of a venue among all users. Moreover, we considered only venues in the Arts, Education, Food, Nightlife, and Work categories, which we expect to better capture differences in gender preferences for venues in a society. We discarded categories that have many subcategories with expected biases towards a particular gender (e.g., Men's Store) as well as categories covering places that might be more popular among non-locals (e.g., hotels and airports), as our goal is to identify gender patterns among residents of particular regions.

Furthermore, when analyzing a particular region,  we only considered  venues of a given subcategory if there are at least two different venues of that subcategory meeting the aforementioned filter criteria in the given region. Finally, we selected 15 countries covering different regions of the world: Brazil, Mexico, and United States (America); France, Germany, Spain, and United Kingdom (Europe); Japan, Malaysia, South Korea, and Thailand (East and South Asia); Kuwait, Saudi Arabia, Turkey, and United Emirates Arab (Western and Middle-East Asia).
\willi{To ease the computational effort we kept the number of check-ins per country below 30,000 by randomly sampling check-ins belonging to a fixed number of venues. This step was only necessary for Turkey and Malaysia.}

\willi{The filtered dataset, which is used in our analyses, contains a total of $170,665$ check-ins performed by $118,902$ users in $14,982$ venues, distributed across $15$ countries, as detailed in  Table \ref{tab:dataset}. We note that male users account for at least half of all check-ins in $10$ of the selected countries. The number of subcategories that passed in our filtering criteria for each country are: $126$ for Brazil; $9$ for France; $12$ for Germany; $74$ for Japan; $34$ for Kuwait; $116$ for Malaysia; $129$ for Mexico; $38$ for Saudi Arabia; $11$ for South Korea; $15$ for Spain; $85$ for Thailand; $95$ for Turkey; $8$ for the United Arab Emirates; $28$ for the United Kingdom; and $120$ for the United States.}

\willi{
\begin{table}[t!]
 \centering
 \small
  \begin{tabular}{  l | r  | r | r   }
      Country        & Check-ins (\% By Male Users) & Venues  & Users    (\% Male)  \\ \hline \hline
       Brazil        & 29,017    (49\%)             &  3,042  &  20,164  (49\% male)  \\ \hline
       France        &   422     (60\%)             &    38   &     337  (61\% male)   \\ \hline
      Germany        &   329     (76\%)             &    35   &     309  (77\% male) \\ \hline
        Japan        & 12,326    (86\%)             &  1,028  &   7,919  (85\% male)  \\ \hline
       Kuwait        &  3,816    (45\%)             &   243   &   2,308  (45\% male)  \\ \hline
     Malaysia        & 29,599    (56\%)             &  2,685  &  17,101  (54\% male)  \\ \hline
       Mexico        & 29,963    (59\%)             &  2,892  &  19,660  (59\% male) \\ \hline
 Saudi Arabia        &  3,576    (39\%)             &   342   &   2,714  (39\% male)  \\ \hline
  South Korea        &   297     (39\%)             &    33   &     250  (42\% male)   \\ \hline
        Spain        &   467     (74\%)             &    58   &     432  (74\% male)  \\ \hline
     Thailand        & 14,579    (23\%)             &  1,346  &   8,772  (23\% male) \\ \hline
United Arab Emirates &   211     (55\%)             &    27   &     187  (56\% male)  \\ \hline
United Kingdom       &  1,061    (69\%)             &   115   &     920  (70\% male)   \\ \hline
United States        & 15,633    (60\%)             &  1,756  &  11,686  (61\% male) \\ \hline
       Turkey        & 29,369    (54\%)             &  1,470  &  26,336  (53\% male)  \\
    \end{tabular}
    \caption{Overview of our dataset.}
    \label{tab:dataset}
    \vspace{-4mm}
\end{table}
}

\section{Characterization of Cultural Gender Preferences for Venues}\label{secCharacterization}

In this section, we present our   methodology to analyze gender preferences for venues   in different regions  around the world, which are known to present some cultural differences \cite{inglehart:2010}. We start by  introducing our methodology (Section \ref{subsec:meth}), and then illustrate how it is applied to study gender preferences for venues at  the country level (Section \ref{subsec:country_analysis}) and at finer granularities (Section \ref{subsec:smallerregions_analysis}).

\subsection{Proposed Methodology} \label{subsec:meth}
\subsubsection{Estimating Gender Preferences} \label{subsec:meth:preferences}
The first step  in our methodology is to  characterize the preferences within each gender group  for different locations in a given region. To that end, we extract  check-ins in  venues  located in the region under study from Foursquare and use them to map the preferences of each gender for specific venues  in the region.
Our methodology is general enough to consider all venues of the same type (same subcategory)  jointly,
or each venue individually, depending on the goal of the study. In the following description, we consider the former, but in Section \ref{subsec:smallerregions_analysis}  we show how it can be easily applied to study cross-gender differences in preferences for individual venues.

Given each venue subcategory that passed our filtering criteria in the region under study (Section \ref{secDatasetUsed}),  we  measure the popularity of all venues of that subcategory within each gender group.  That is,  given a region, a subcategory, a venue, and a gender, we compute the percentage of all check-ins  by users of that gender in all venues of that region that were performed in venues of the given subcategory. \willi{To make the graphs better comparable}, we normalize these percentages by \willi{dividing by} the maximum value, only to ease the visualization.

The next step consists in computing the cross-gender popularity difference $d_s$ for each subcategory. Let us define a 2-dimensional space based on the two popularity measures (one per gender). The diagonal of this space represents an ideal case where popularity is balanced across genders. The cross-gender popularity difference for a given subcategory is then defined  as the  shortest euclidean distance between the point representing that particular subcategory in the 2-dimensional space  and its diagonal\footnote{We did experiment with other approaches to computing the popularity difference, such as the difference between the coordinates but the results are similar.}. \willi{Differences below zero indicate greater popularity among female users as the point lies on the left side of the diagonal. In contrast, differences above zero imply greater popularity among male users.}

Given a non-zero cross-gender popularity difference, computed as described, a natural question that emerges   is: Is this difference related to a possible difference  in size of the female/male population in the studied dataset, or does it reflect a significant gender-related pattern?

\begin{figure*}[!hbt]
  \centering
  \subfigure[Brazil (observed)]
              {\includegraphics[width=.23\textwidth]{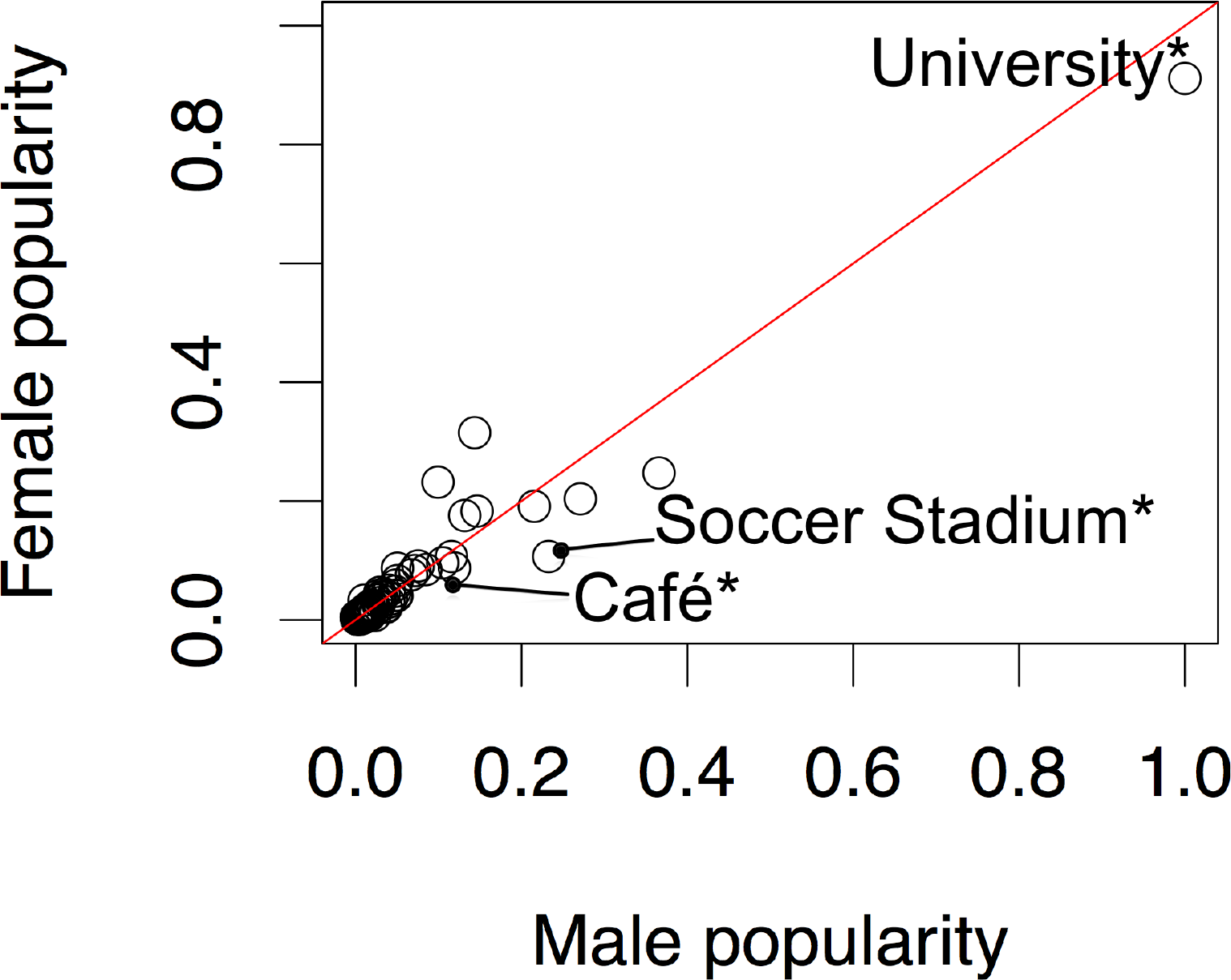}}
    \subfigure[Brazil (null model)]
              {\includegraphics[width=.23\textwidth]{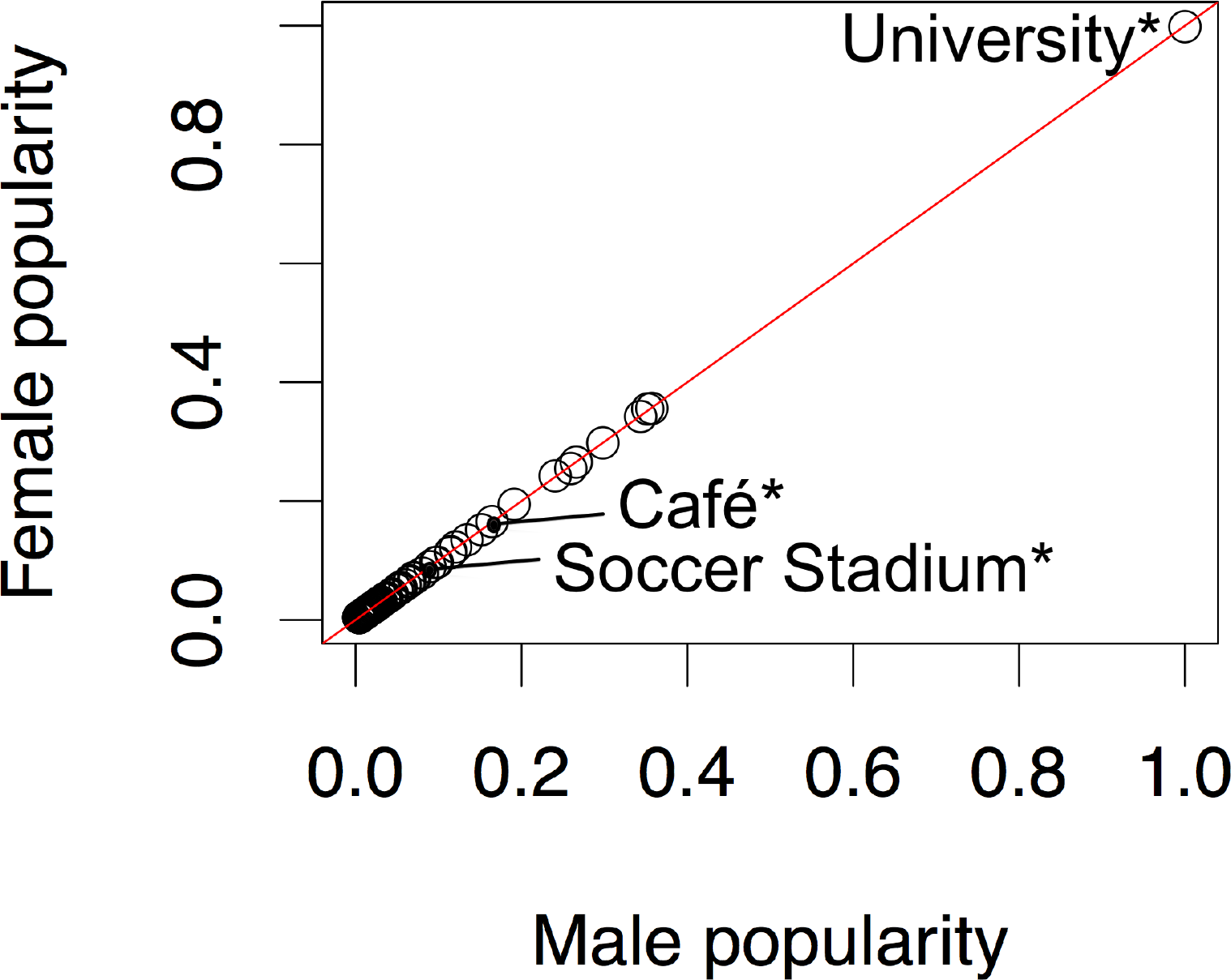}}
    \subfigure[USA (observed)]
              {\includegraphics[width=.23\textwidth]{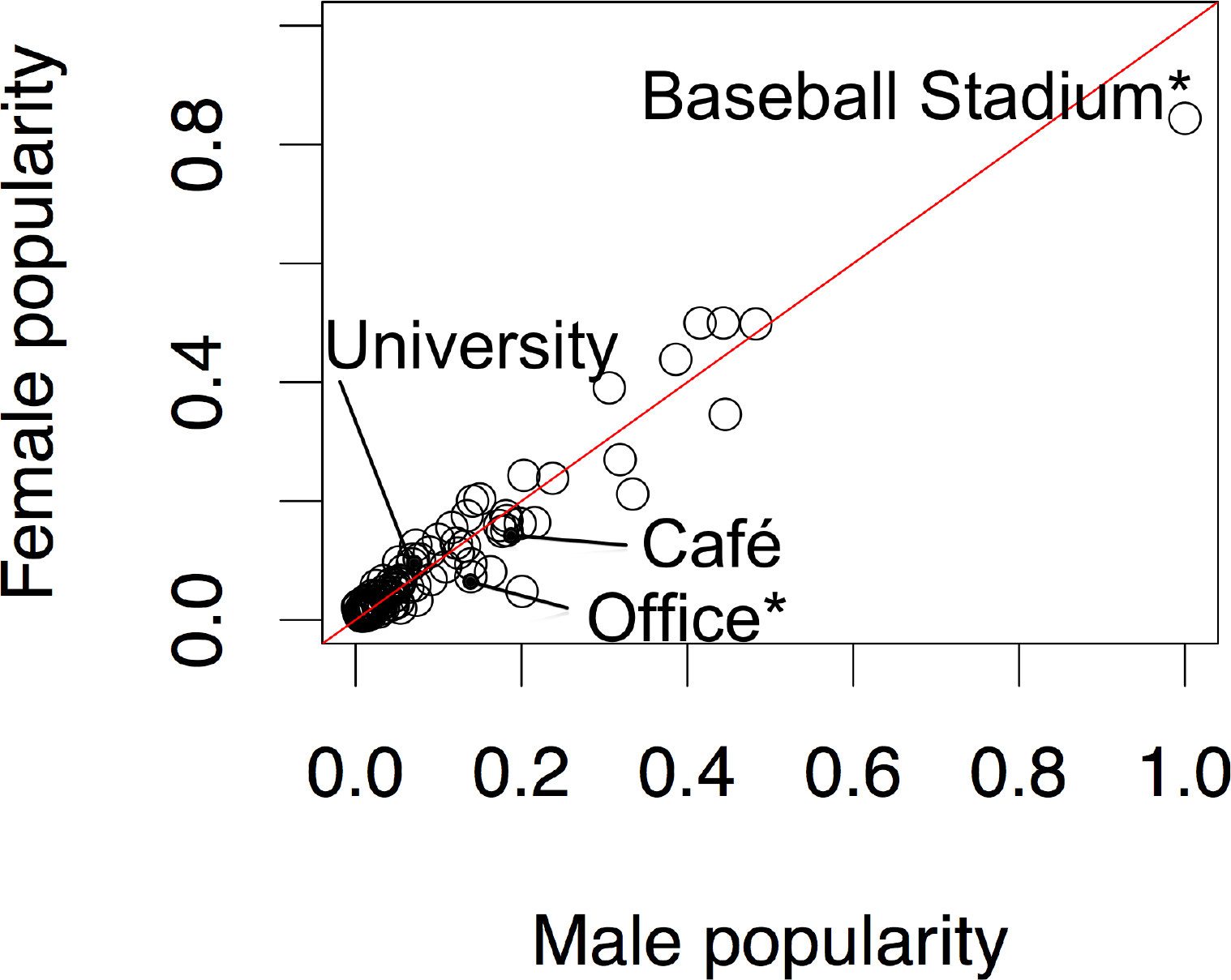}}
    \subfigure[USA (null model)]
              {\includegraphics[width=.23\textwidth]{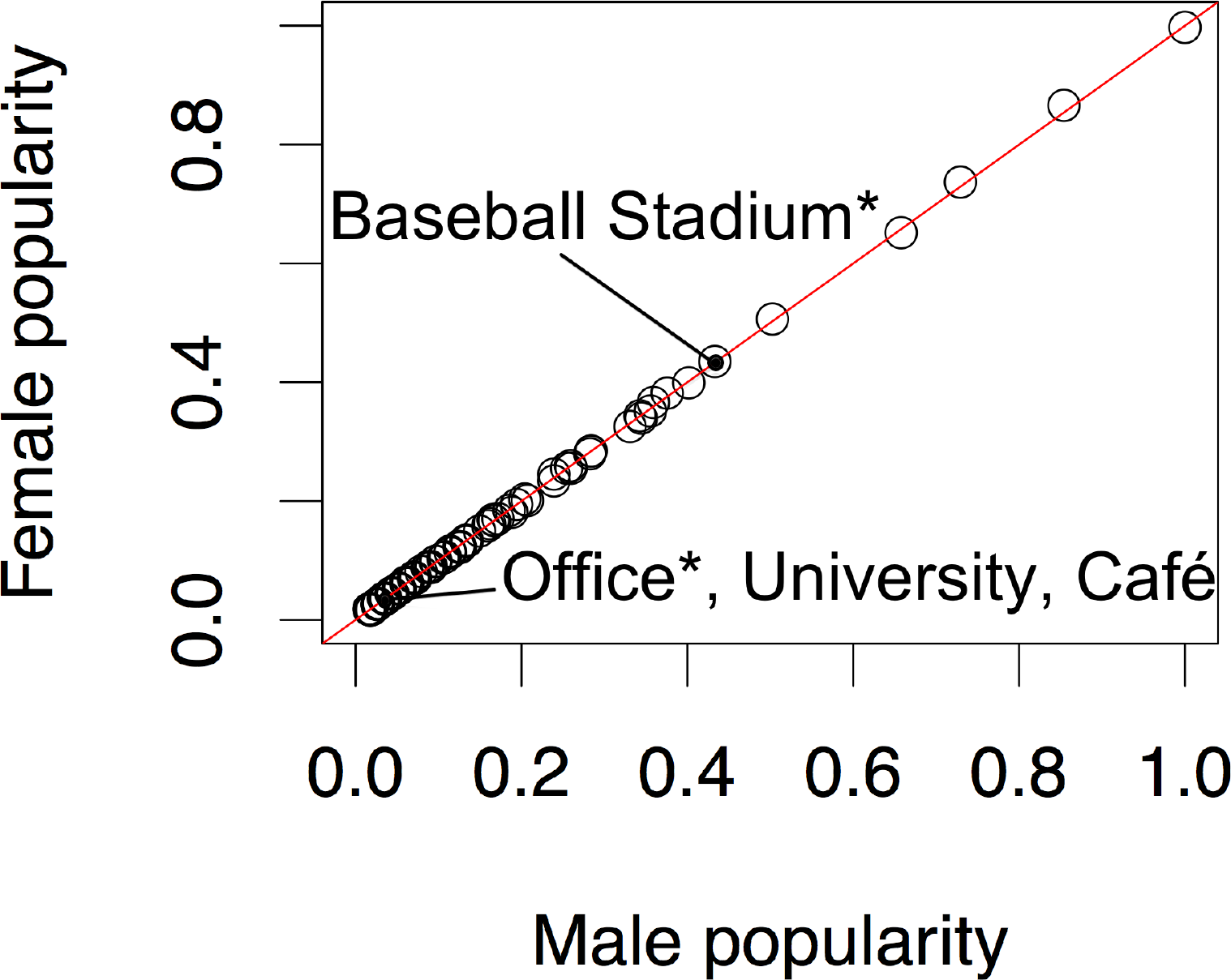}}
 \caption{Popularity (normalized) of venue subcategories within each gender for Brazil and United States, and the average values after a null model creation for the same country.}\label{fig:segregation-subc-countries}
 \end{figure*}

\begin{figure*}[!hbt]
  \centering
    \includegraphics[width=.99\textwidth]{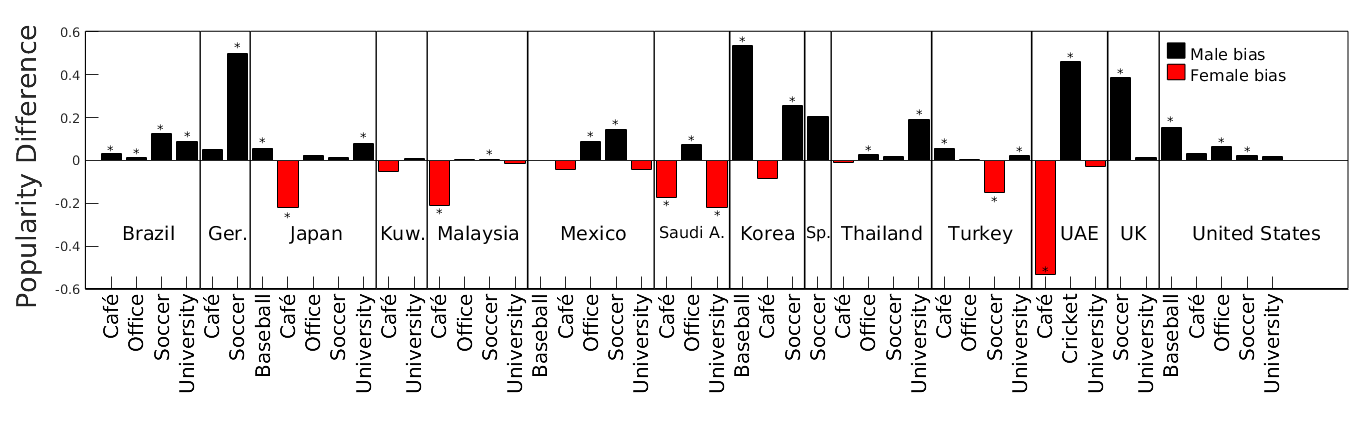}
 \caption{\thiago{Popularity difference of venue subcategories within each gender in various countries. For each country we show the subcategories {\it Baseball Stadium}, {\it Café}, {\it Cricket Ground}, {\it Office}, {\it Soccer Stadium}, and {\it University}. The differences represent normalized values for each country, to facilitate the comparison.}}\label{fig:subc-countriesAllInOne}
 \end{figure*}

\subsubsection{Testing Statistical Significance} \label{sec:testStat}

To tackle this question, we built a null model using the following process: \thiago{We count the number $c$ of  all check-ins located in the region under study. Furthermore, we group all unique users in $U$ and all locations in $L$ (preserving the venue's attributes, i.e., subcategory, latitude, and longitude). After that, we generate $c$ check-ins randomly choosing for each of them a gender (female or male), a location in L, and a user in U. Any element (gender, location, or user) is randomly sampled with replacement and thus can be chosen more than once. In this way, we disjoint the correlation between the user, gender, and location. We then recompute the cross-gender popularity difference for each subcategory as discussed in Section \ref{subsec:meth:preferences}.

We repeat this process $k$=$100$ times, producing a distribution  of popularity differences for each subcategory. By comparing the observed difference for a given subcategory against the corresponding distribution produced by the aforementioned randomization process, we are able to rule out any possible effect due to differences in gender population sizes. Also, we can test whether the observed cross-gender difference is significant, meaning that it is indeed related to gender preferences.

Let $d_s$ be the observed difference for subcategory $s$, and $D^{null}_s$ the distribution of differences obtained after randomization. We compare $d_s$ against $D^{null}_s$ with the minimum $min$ and maximum $max$ limits representing the values observed in $D^{null}_s$ with 99\% of confidence.  The observed difference is significant if it lies {\it outside} the range [$min, max$].  We refer to the range of values against which  $d_s$ is tested as the {\it acceptance range} [$\Delta_{min},\Delta_{max}$].  If $d_s$ lies inside this range, it cannot be considered significant, and we cannot tell whether it actually reflects a gender-related pattern. }

\thiago{We also tried another randomization approach, preserving all check-in attributes unchanged, except gender,  and randomly shuffling $k=100$ times the gender associated with all check-ins located in the region under study. Yet, the results are similar to the discussed above. For this reason, in this study, we only present more details and discuss results of the approach mentioned previously. Next, we illustrate the use of our methodology in various scenarios.}

\subsection{Country-Level Analysis} \label{subsec:country_analysis}

We start by focusing on a coarser spatial granularity and use our methodology to analyze gender preferences for venue subcategories across different countries. Figure \ref{fig:segregation-subc-countries}\footnote{In this figure and also in Figures \ref{fig:subc-countriesAllInOne} and \ref{figGenderSegSP} \say{$*$} means that the difference observed is statistically significant.} shows  the (normalized) popularity, within male and female users, of  considered  subcategories  in  Brazil (Figure \ref{fig:segregation-subc-countries}a) and United States (Figure \ref{fig:segregation-subc-countries}c). Each point in each graph represents a subcategory, which only some examples are labeled to avoid visual pollution. In Figures \ref{fig:segregation-subc-countries}a and c soccer and baseball stadiums are the most popular subcategories, respectively, both biased towards male users.

\thiago{We analyzed all subcategories that passed our filtering criteria in each country, but we here discuss only some of the most popular examples in terms of the number of check-ins: Baseball Stadium, Café, Cricket Ground, Office, Soccer Stadium, and University. Figure \ref{fig:subc-countriesAllInOne} shows the popularity difference of venue subcategories within each gender in all studied countries. To ease the comparison, the differences represent normalized  values (into the range $[0,1]$) for each country. Note, that differences below zero indicate greater popularity among female users, while differences above zero indicate greater popularity among male users.}

Studying the results in Figure \ref{fig:subc-countriesAllInOne}, we can see, for instance, that {\it Soccer Stadiums}, tend to be more popular among male users in all countries except in Turkey. In contrast, {\it Universities} are more popular among male users in Brazil, but more female-oriented in Saudi Arabia. Similarly, there is a cross-gender difference towards men for {\it Cafes} in Turkey and the USA, whereas, in Malaysia and Saudi Arabia, those places tend to attract more female users. Do these differences reflect different gender preferences  in those countries?

We then turn to the results produced after the randomization process, shown in Figure \ref{fig:segregation-subc-countries} (b and d), which presents average popularity values computed across all $k=100$ repetitions.  Note that, unlike  in the observed data, those values are well balanced across genders in all cases. \thiago{This pattern repeats for all studied regions, for this reason, we only show two illustrative examples.}

\begin{figure*}[!hbt]
  \centering
    \subfigure[Soccer Stadium, Brazil]
              {\includegraphics[width=.20\textwidth]{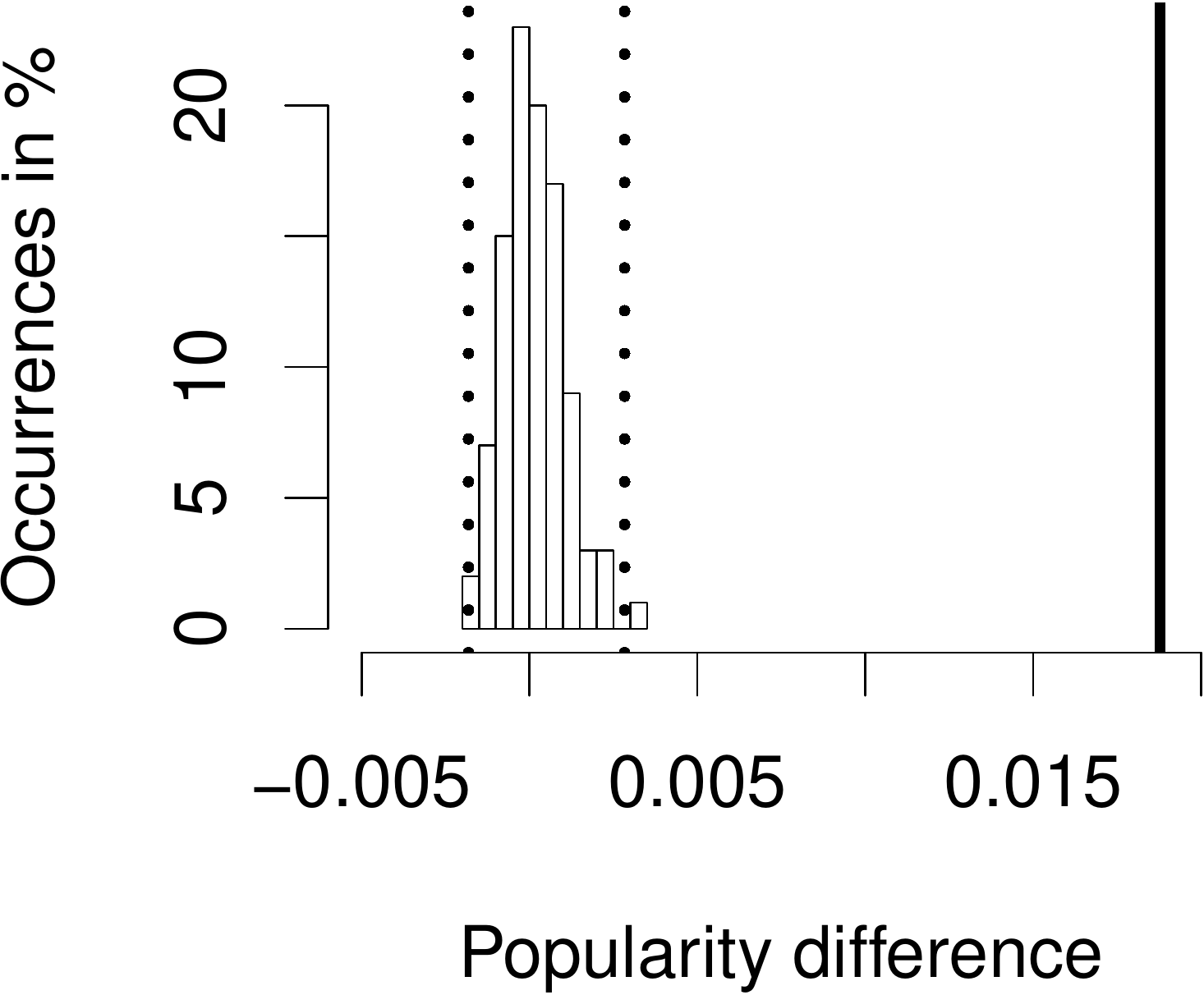}}
    \subfigure[Soccer Stadium, Turkey]
              {\includegraphics[width=.20\textwidth]{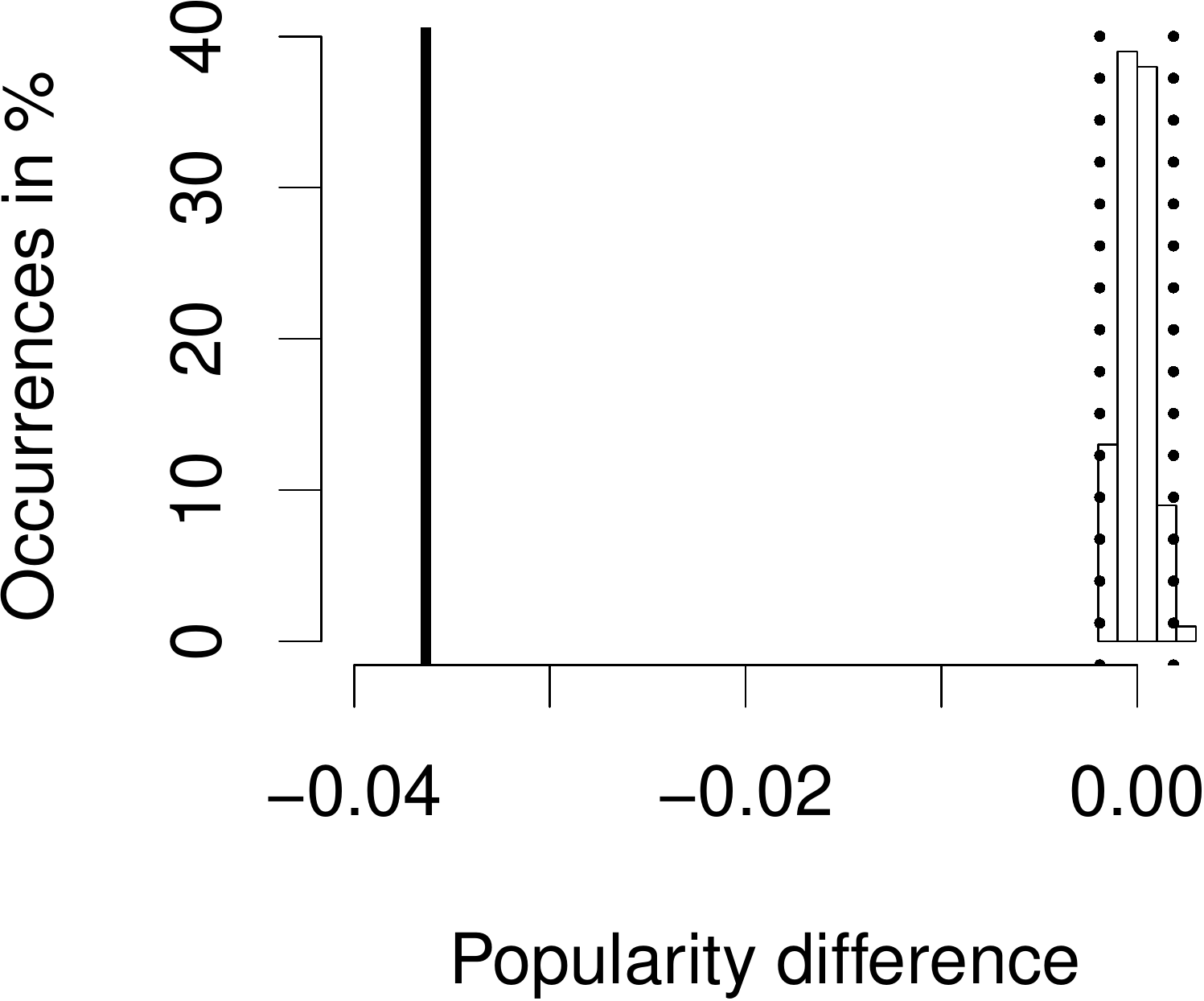}}
 \subfigure[Baseball Stadium, USA]
              {\includegraphics[width=.20\textwidth]{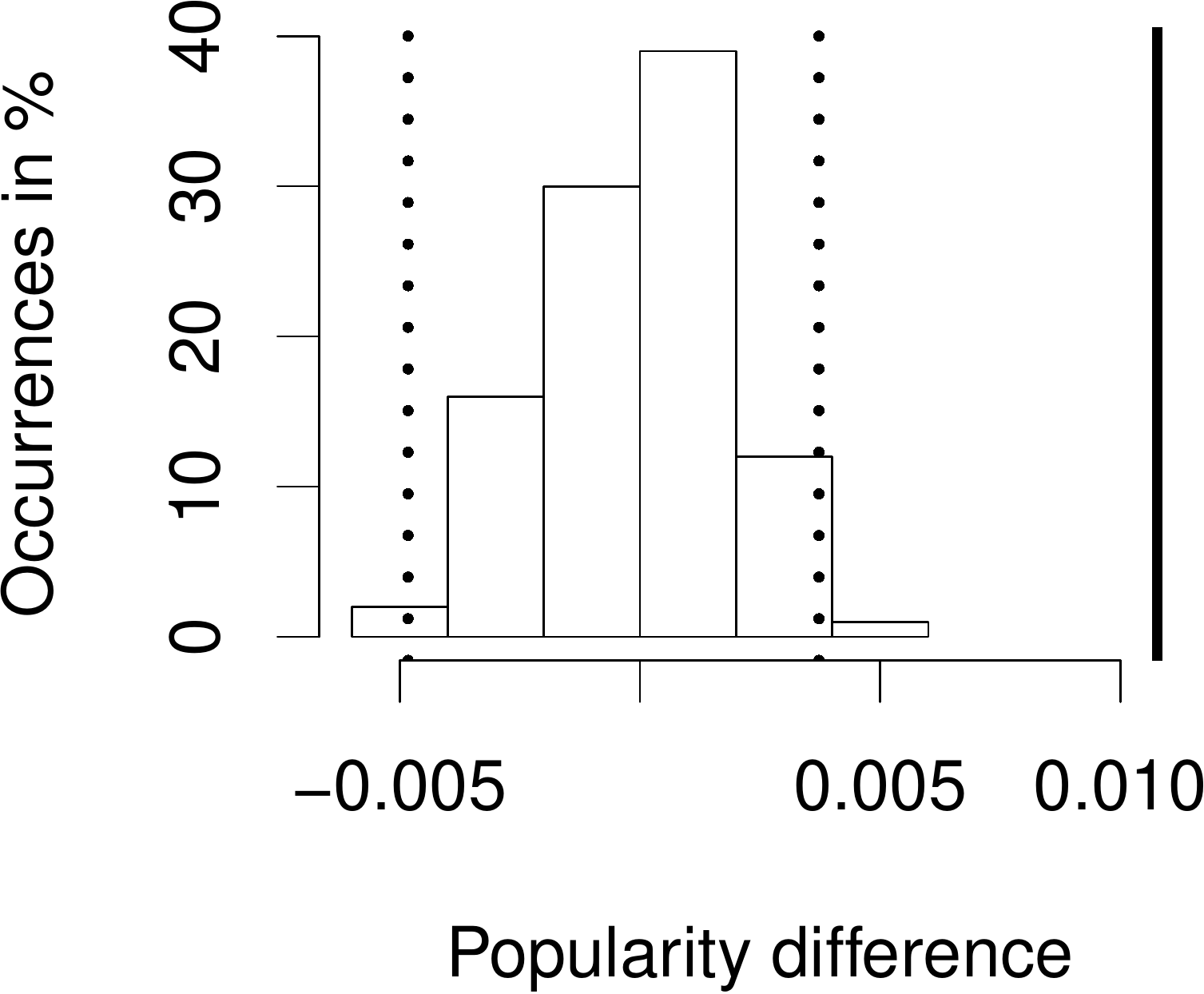}}
    \subfigure[Baseball Stadium, Mexico]
              {\includegraphics[width=.20\textwidth]{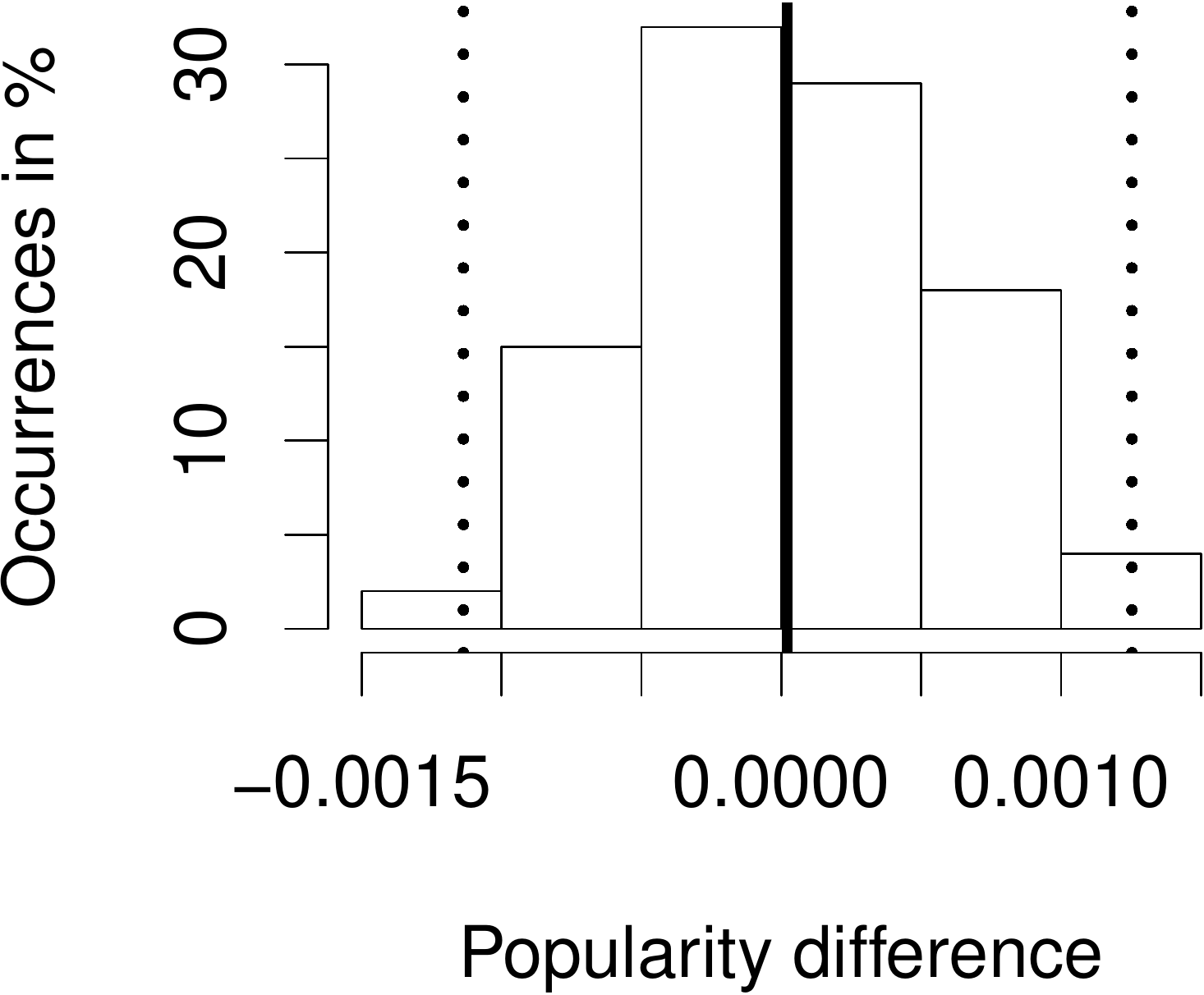}}
 \subfigure[Cricket Ground, UAE]
              {\includegraphics[width=.20\textwidth]{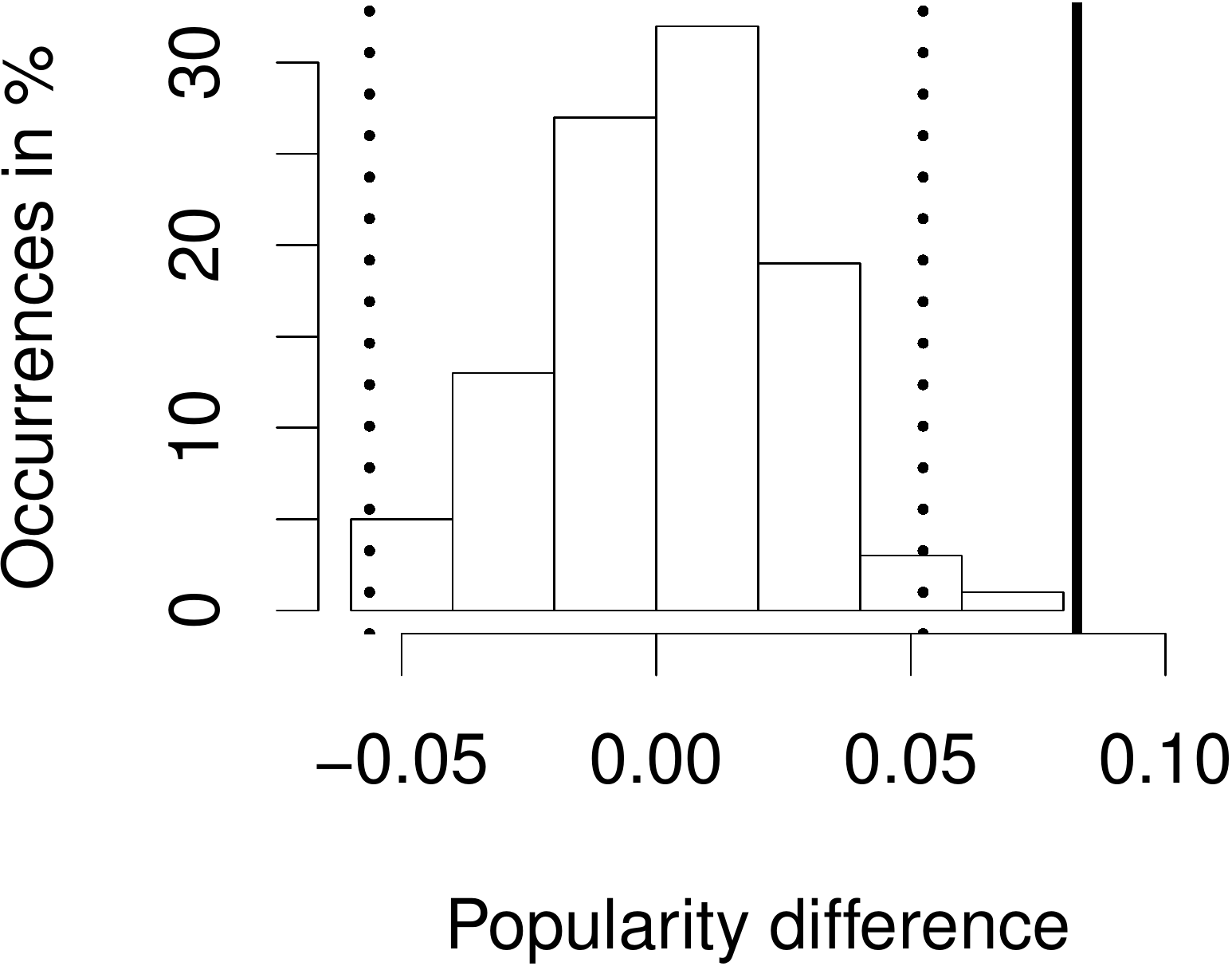}}
 \subfigure[Café, Japan]
              {\includegraphics[width=.20\textwidth]{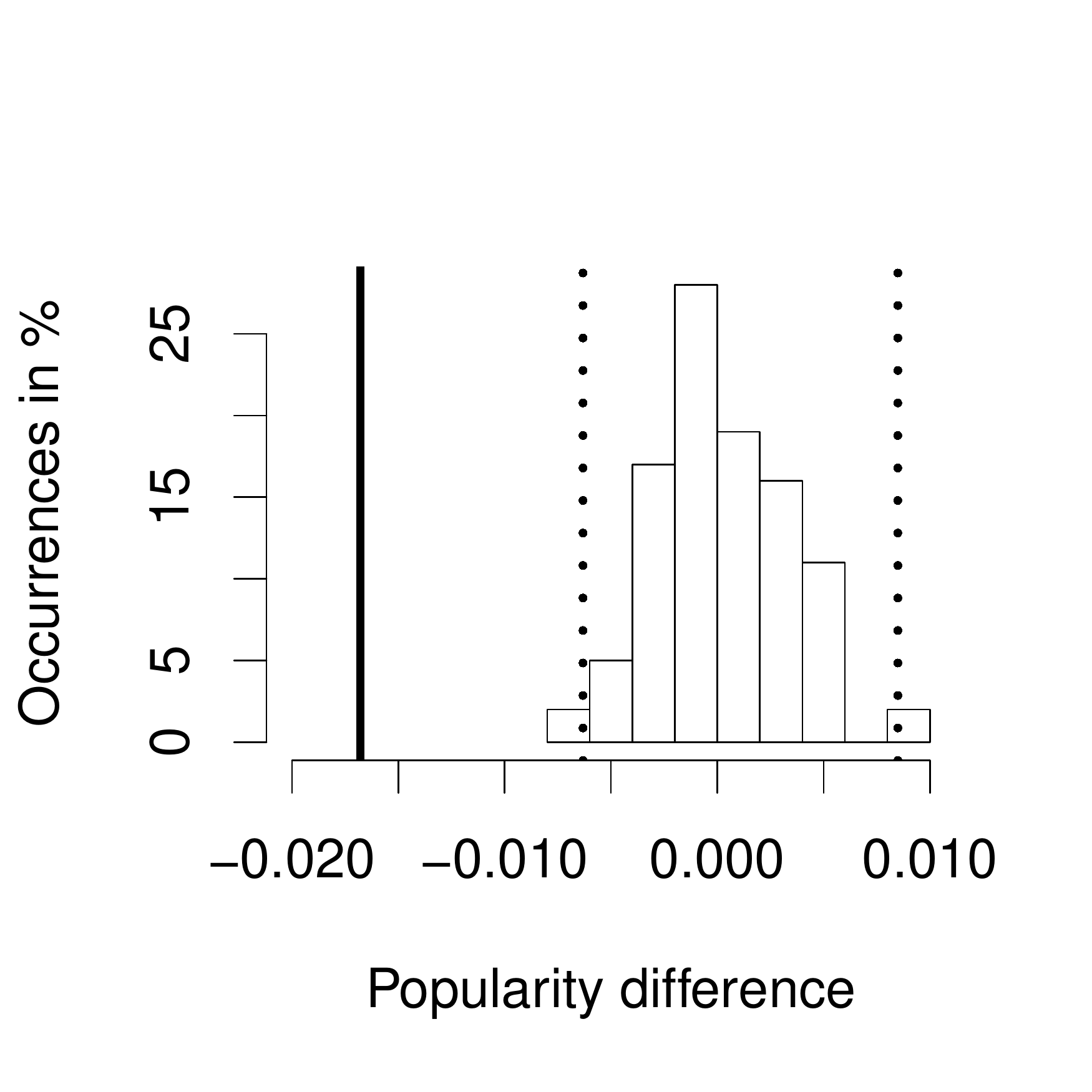}}
 \subfigure[Café, Brazil]
              {\includegraphics[width=.20\textwidth]{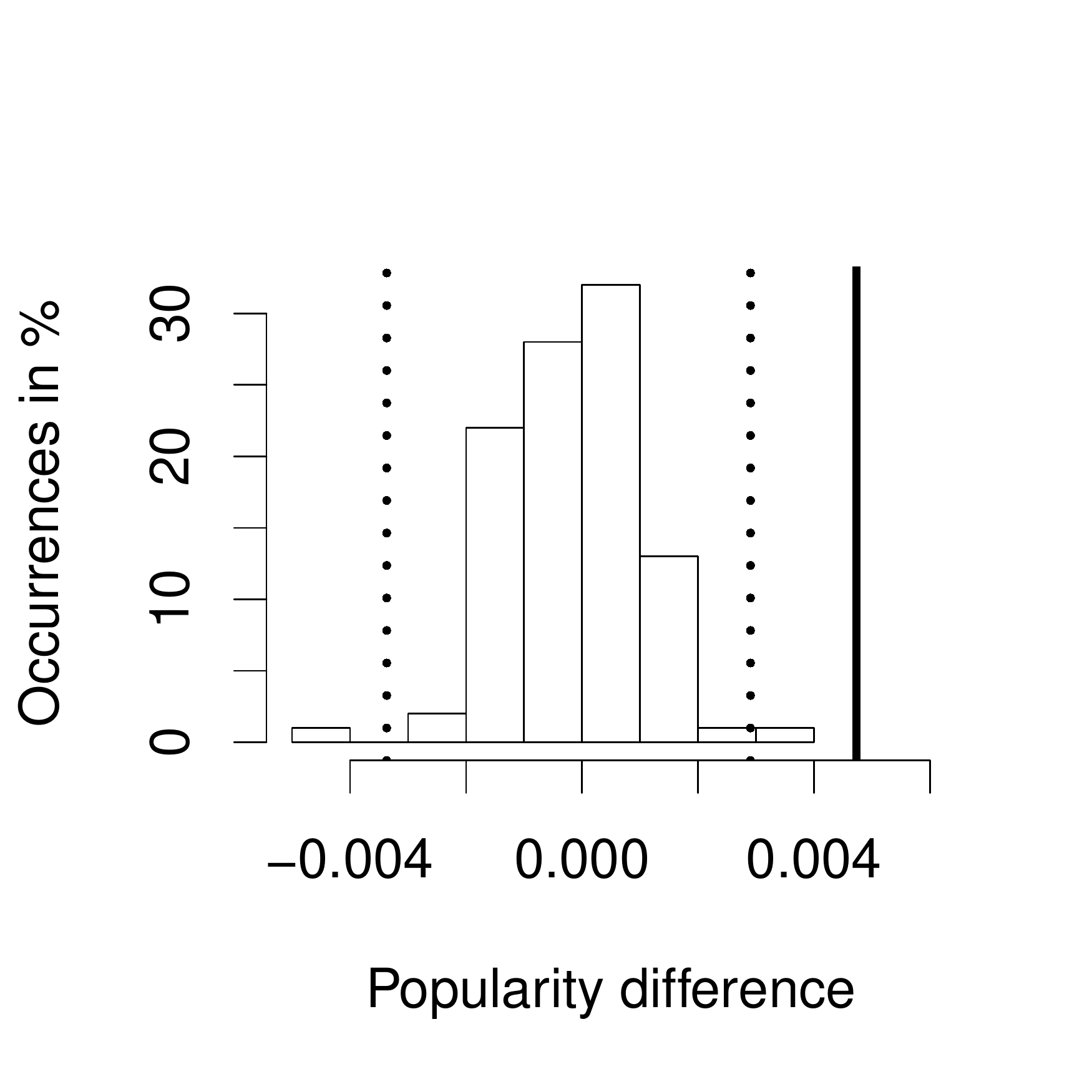}}
 \subfigure[University, Saudi Arabia]
              {\includegraphics[width=.20\textwidth]{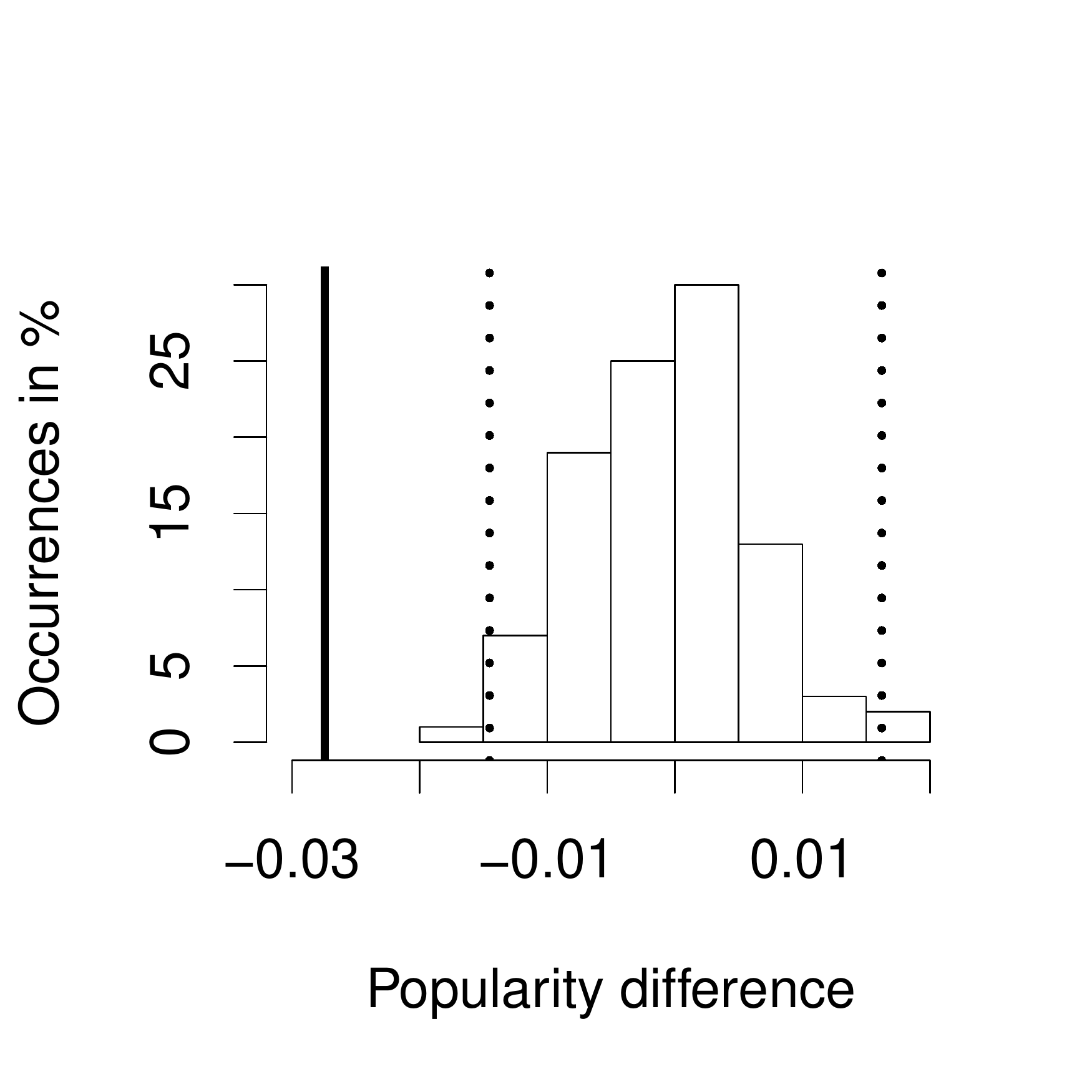}}
              \caption{Distribution of cross-gender popularity differences produced by randomization process for various subcategories and countries. The dashed lines mark the acceptance range [$\Delta_l$,  $\Delta_u$], and the solid line the observed value $d_s$. Figures (a,c,e,f,g,h) show significant cross-gender differences, whereas (b,d) do not.}
\label{fig:histogramCountry}
\vspace{-4mm}
\end{figure*}

We delve further into some of the  results shown in Figure \ref{fig:segregation-subc-countries},
starting with  three particular subcategories related to sports, namely {\it Soccer Stadium}, {\it Baseball Stadium}, and {\it Cricket Ground}. Out of all analyzed countries, we find that {\it Soccer Stadiums} are significantly more popular among male users, i.e. have statistically significant cross-gender differences above zero in Brazil, Mexico, Germany, \willi{South Korea, the USA,} Malaysia and the UK. As an example, Figure \ref{fig:histogramCountry}a shows the distribution of the cross-gender differences computed during the randomization procedure for Brazil. The solid vertical line is the   difference observed in the data ($d_s$), whereas the dashed vertical lines indicate the acceptance range $[\Delta_{min}, \Delta_{max}]$.  Note that the observed difference ($0.0188$) by far exceed the upper limit $\Delta_{max}$.

In contrast, in Spain, Japan, and Thailand, the cross-gender popularity differences were not significant, according to our test. This might be due to a greater popularity of the female soccer  teams in these countries, which attract proportionally more male users to related venues, compared to Brazil, Mexico and the other aforementioned countries. Turkey, however, is an interesting case: We found a difference significantly below zero, indicating a \willi{far higher preference among female users, result shown in Figure \ref{fig:histogramCountry}b. This is most likely a consequence of a penalty, introduced in 2011, for Turkish soccer clubs that only women and children under 12 years are allowed to attend games of clubs sanctioned because unruly fans}\footnote{https://www.opendemocracy.net/can-europe-make-it/aslan-amani/football-in-turkey-force-for-liberalisation-and-modernity.}. \willi{In fact, 90\% of the $2,536$ check-ins performed in Turkish soccer stadiums in our dataset were performed in the stadium of {\it Fenerbace Istanbul}. This club was affected by that penalty,  being obligated to ban male teenagers and adults of its stadium during our collection period. During this period this club hosted a game over $50,000$ spectators\footnote{http://www.dailymail.co.uk/sport/football/article-2614502/Turkish-delight-Fenerbahce-wrap-19th-league-title-win-50-000-women-children.html}}.

Turning our attention to the  {\it Baseball Stadium} subcategory, we find that those venues are significantly more popular among male users in \willi{Japan,} South Korea and the USA. The distribution of the cross-gender differences computed during the randomization procedure for this subcategory for the USA is shown in Figure \ref{fig:histogramCountry}c. In contrast, in Mexico, we find no significant trend towards any gender, as shown in Figure \ref{fig:histogramCountry}d.

The {\it Cricket Ground} subcategory was only analyzed for the United Arab Emirates (UAE), as venues in this subcategory in the other countries did not pass our filtering criteria. For that country, where this subcategory was the most popular type of sports-related venue, we did find a statistically significant positive cross-gender difference, indicating a greater popularity among male users (Figure \ref{fig:histogramCountry}e). Interestingly, a general result for all three sports subcategories is that the overall most popular subcategory of sports venues in the country is often significantly more male-oriented.

Regarding other venue subcategories, we find that {\it Offices} are significantly more popular among male users in all countries \willi{ with sufficient data about this subcategory,} but \willi{Turkey,} Japan, and Malaysia. In the case of Malaysia, the exception might be due to the fact that most popular venues classified as \emph{Office} are also located in shopping malls, which traditionally attract many women, thus compensating for any possible male bias. This also happens in Japan, and  besides that, among the most popular offices there is a Korean-pop record label, a style that has a mostly female audience\footnote{http://www.theguardian.com/music/2011/dec/15/cowell-pop-k-pop.}, indicating that this office may attract many female fans.

{\it Cafes}, in turn, only have a significant cross-gender popularity difference in \willi{$6$} out of \willi{$9$} analyzed countries with sufficient data about cafes. While these places are female-oriented in Japan, Malaysia, \willi{Saudi Arabia}, and the United Arab Emirates, they are more popular among male users in Brazil and Turkey. One possible reason that helps to explain this result is that most popular {\it Cafes} analyzed in Brazil are located in popular areas among men, such as offices and financial regions. In Turkey, it is usually men who most frequent cafes, although these also now welcome women \cite{turkeycoffee}. We illustrate this finding by presenting the results for Japan and Brazil in Figures \ref{fig:histogramCountry}f and \ref{fig:histogramCountry}g, respectively. These results  illustrate significantly different cross-gender patterns in both countries.

As a final example,  the subcategory {\it University} is significantly more popular among male users in Brazil, Japan, Thailand\willi{, and Turkey} but, as shown in Figure \ref{fig:histogramCountry}h, much more female-oriented, with significant differences, in Saudi Arabia. One possible explanation for the latter is that the majority of university graduates are women in Saudi Arabia, according to a recent report\footnote{http://www.worldpolicy.org/blog/2011/10/18/higher-education-path-progress-saudi-women.}.

Our goal in this section was to illustrate the use of the proposed methodology to characterize gender preferences for different types of locations in a country. As discussed above, our results do suggest that the observed differences reflect inherent cultural aspects of each country.

\subsection{Finer Grained Analyses} \label{subsec:smallerregions_analysis}

In the previous section, we showed how our methodology  can be used to identify significant cross-gender differences in preferences for venues in different countries. We now show that it can also help identify  such differences at much finer granularities. Focusing on a specific city -- S\~ao Paulo (Brazil) -- we study differences in gender preferences for  specific venues in two scenarios: all venues in the city, and all venues of a given subcategory. The latter is useful to identify places where gender preferences patterns diverge from those of the same type in the city.

\begin{figure}[!ttt]
  \centering
    \subfigure[All venues]
              {\includegraphics[width=.35\textwidth]{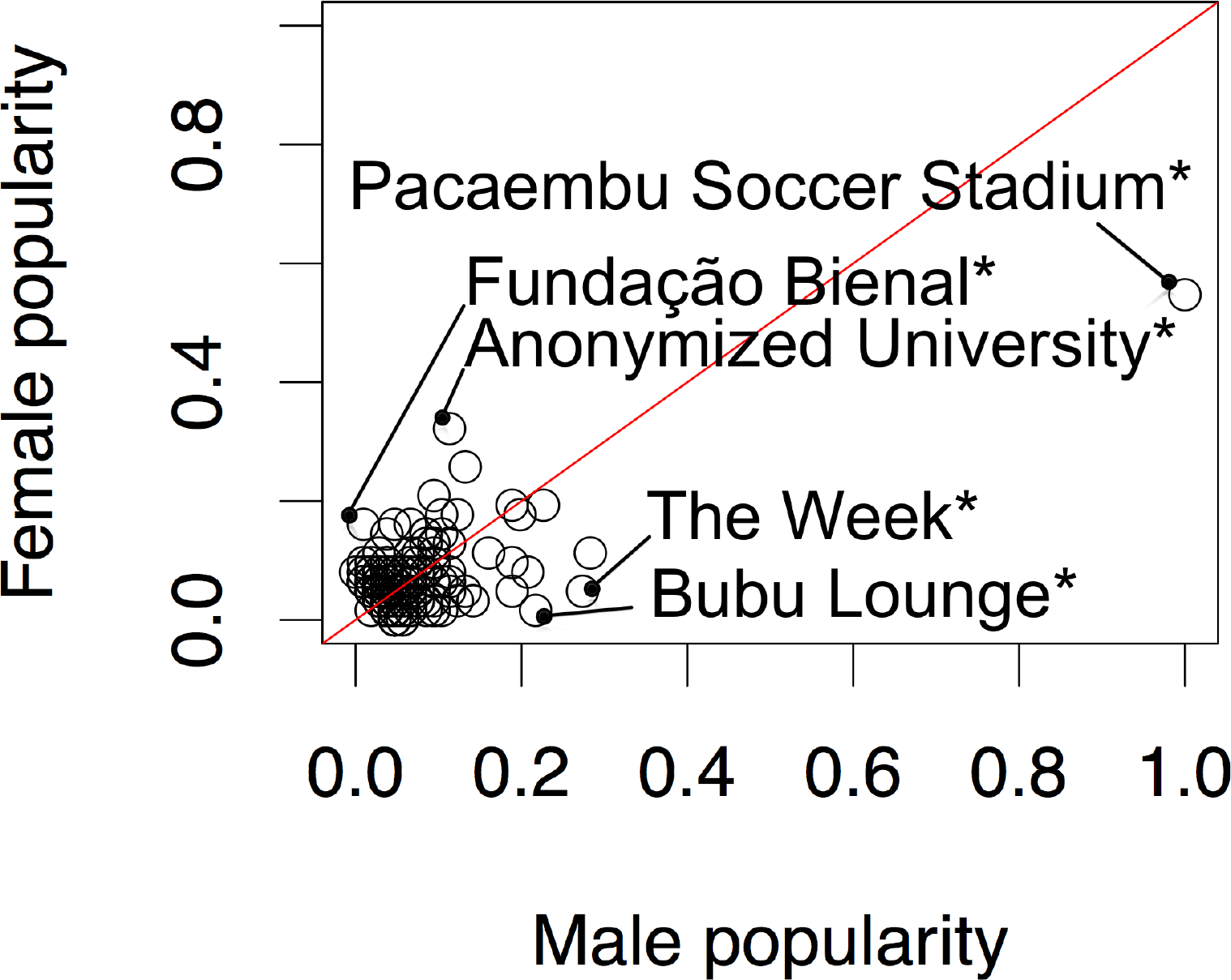}}
    \subfigure[Nightclub venues]
               {\includegraphics[width=.35\textwidth,height=100pt]{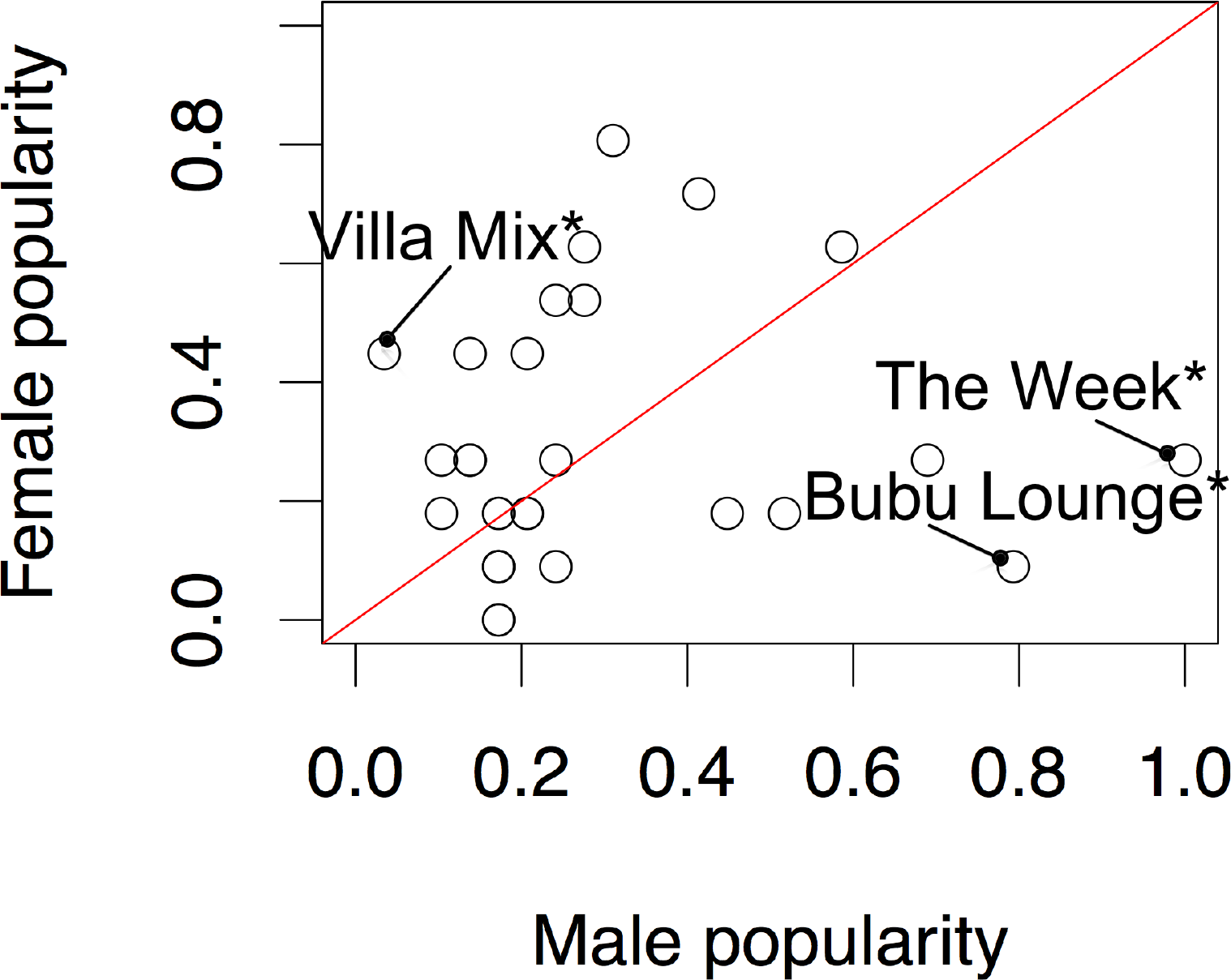}}
\caption{Popularity (normalized) of individual venues within each gender group in S\~{a}o Paulo, Brazil (left: all values from all subcetegories; right: only venues from the subcategory Nightclub).}
\label{figGenderSegSP}
\vspace{-4mm}
\end{figure}

In the first scenario, we applied our methodology considering \willi{$2,422$} check-ins at   venues located in S\~{a}o Paulo. \thiago{Figure \ref{figGenderSegSP}a shows these results for the observed data \thiago{(normalized just to ease the visual evaluation}).} As Figure \ref{figGenderSegSP}a shows, there are some large cross-gender differences in the city. Out of all \willi{$248$} venues analyzed, we identified \willi{$21$} where the cross-gender popularity difference is statistically significant, according to our methodology.

\begin{figure}[!bt]
  \centering

    \subfigure[\willi{Anonymized University Campus}]
              {\includegraphics[width=.35\textwidth]{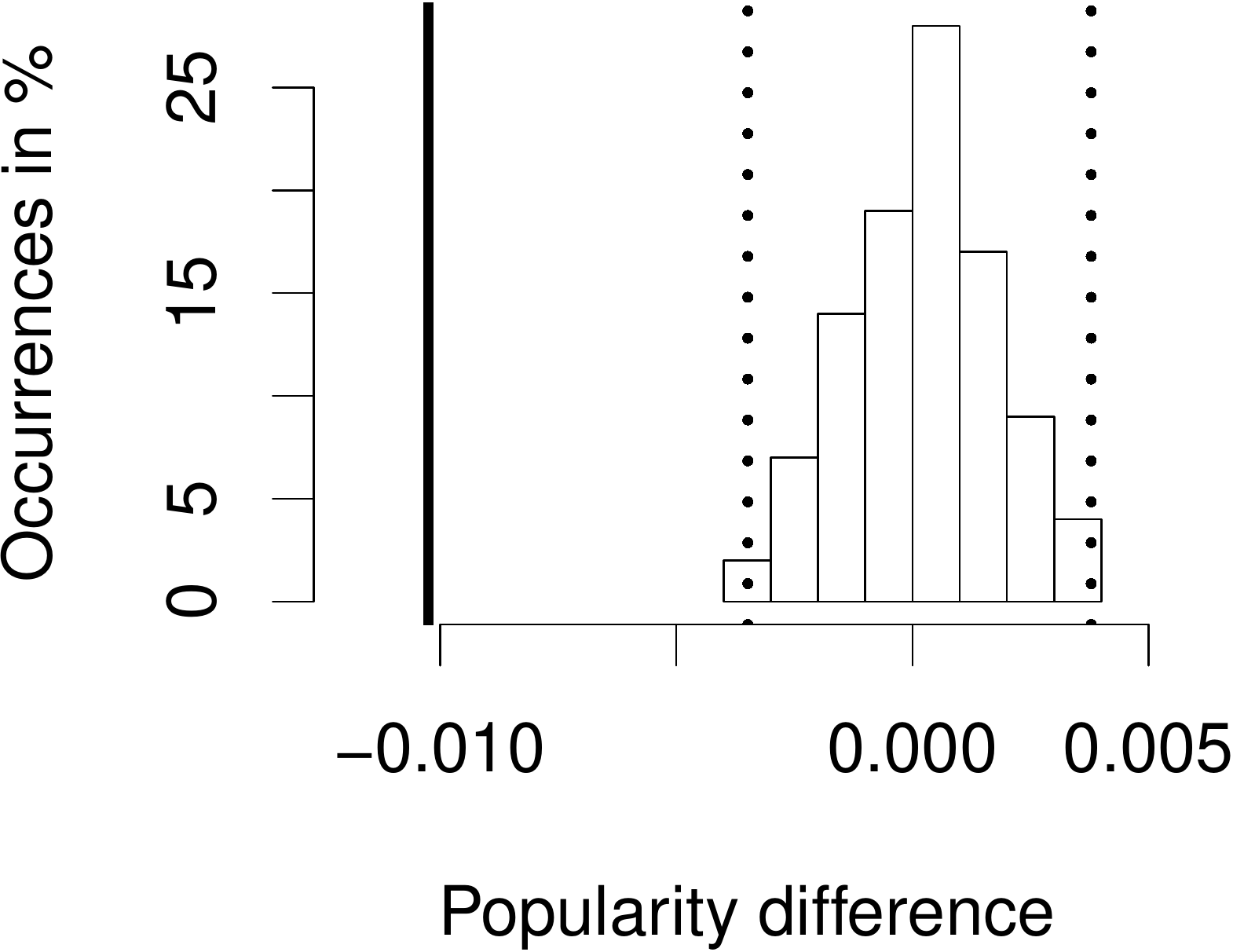}}
    \subfigure[Art Museum Funda\c{c}\~ao Bienal Ibirapuera]
              {\includegraphics[width=.35\textwidth]{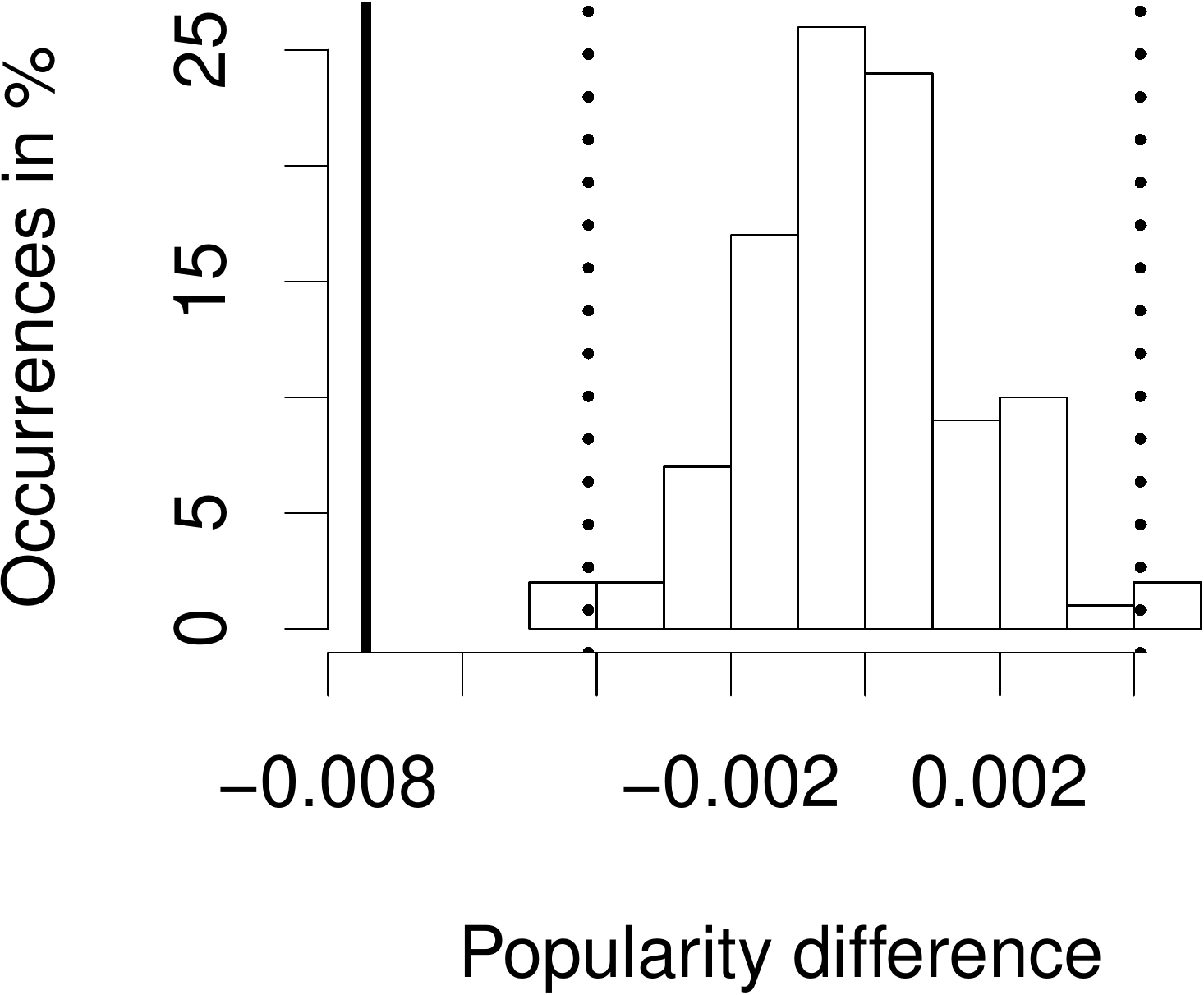}}
\caption{Distribution of cross-gender popularity differences produced by randomization process for two venues in S\~ao Paulo city.}
\label{figGenderSegSPhist}
\vspace{-4mm}
\end{figure}

One such example is a private university, that explicitly requested to be anonymized. It is \willi{more popular among} female users, with a statistically significant cross-gender difference below zero (Figure \ref{figGenderSegSPhist}a). This might be explained by an often larger presence of women in the particular courses located on that campus (namely health, arts, pedagogy, and media production) in Brazil. Similarly, the {\it Technology and Communications University FAPCOM},  which offers similar and related courses, is also significantly  more popular among female users. \willi{A spokesperson for the anonymized university confirmed via email that they indeed have 68\% female students enrolled at the campus our method detected as anomalous.}

Another example is the {\it Art Museum Funda\c{c}\~ao Bienal Ibirapuera}, which is also significantly more popular among female users, as shown in Figure \ref{figGenderSegSPhist}b. This result was confirmed by a spokesperson for this museum. Besides that, the result is consistent with  findings from a recent survey performed with visitors of this museum, confirming that the majority of the public is female \cite{bienal}.

In the second scenario, we considered check-ins at individual {\it Nightclub} venues located in S\~{a}o Paulo. \thiago{To ease the visualization of the results, they were plotted normalized.} As shown in Figure \ref{figGenderSegSP}b, various nightclubs lie far from the diagonal. Yet, out of all $29$ nightclubs analyzed, we found \willi{$4$} with statistically significant cross-gender differences: {\it The Week}, {\it Bubu Lounge}, {\it Villa Mix}, and \textit{Blitz Haus}.

\begin{figure}[!hbt]
  \centering
    \subfigure[The Week]
              {\includegraphics[width=.35\textwidth]{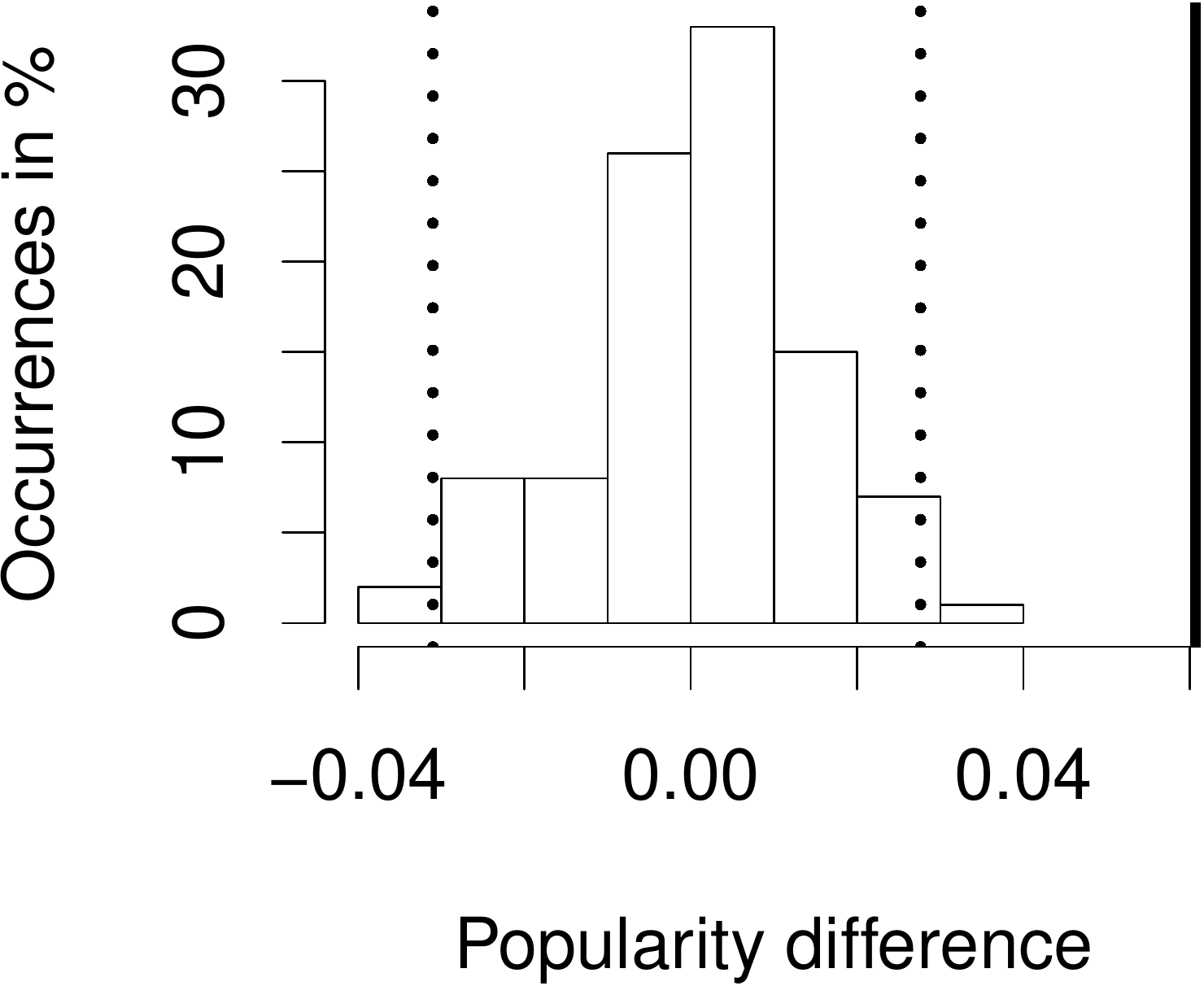}}
    \subfigure[Villa Mix]
              {\includegraphics[width=.35\textwidth]{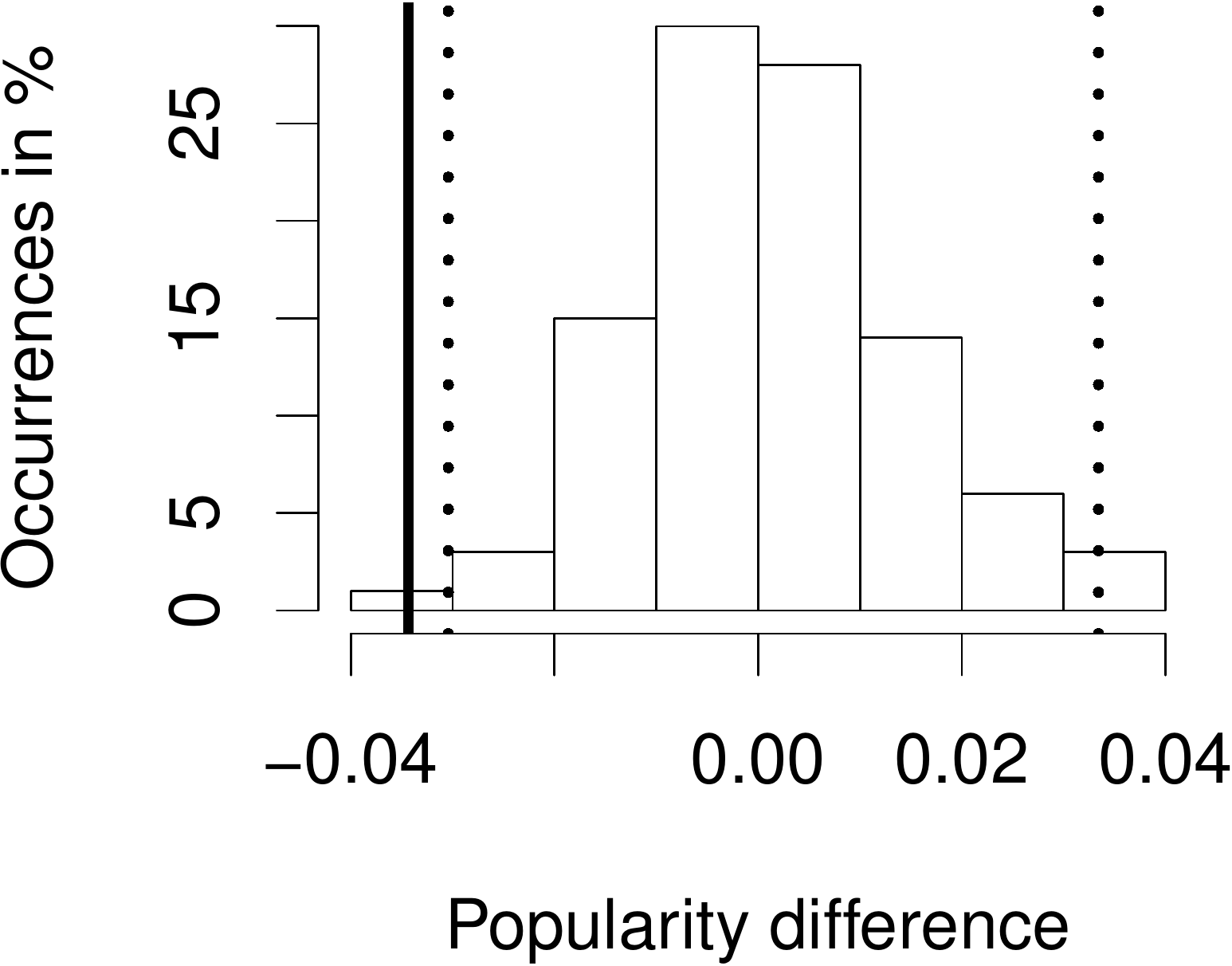}}
\caption{Distribution of cross-gender popularity differences produced by randomization process for two {\it Nightclub} venues in S\~ao Paulo city.}
\label{figGenderSegSPNightclubHist}
\vspace{-4mm}
\end{figure}

{\it The Week} (Figure \ref{figGenderSegSPNightclubHist}a), and \textit{Bubu Lounge}  are significantly more male-oriented. \willi{Supporting our finding, today {\it The Week} and \textit{Bubu Lounge} are classified as a {\it Gay Bar} on Foursquare, which was not the case during our data collection.} \willi{Also, on similar recommendation platforms, such as Yelp\footnote{http://www.yelp.com.}, TripAdvisor\footnote{http://www.tripadvisor.com.} and even specialized ones, such as GayCities\footnote{http://www.gaycities.com.}, they are labeled as \say{gay} and \say{male-dominated}.

In contrast, \textit{Villa Mix}(Figure \ref{figGenderSegSPNightclubHist}b), and \textit{Blitz Haus} are significantly more popular among female users. \willi{The manager of \textit{Villa Mix} confirmed to us via email that they receive more visits of women than men. This might be explained by the fact that this nightclub frequently holds musical events with \textit{Sertanejo} artists, a Brazilian music style that tendd to be popular among Brazilian women. It is important to mention that all venues studied in this section were contacted to confirm our results, and all the replies were mentioned in the text.} For the case of \textit{Blitz Haus} a fact that could help to explain the result is that according to their website\footnote{http://blitzhaus.com.br.}, the nightclub has a retro decoration, and besides music offers a gastronomic place.}

This suggests that our methodology can detect venues that do not follow the same gender preference pattern observed in other venues of the same subcategory in the studied city. This result could be useful, for example, to improve venue classification schemes in the city.

\thiago{Cultural differences, including those related to gender, may exist among different countries~\cite{inglehart:2010,sen2001many,szymanowicz2011intelligent,harrison2000culture}. Besides that, there is a recent evidence that preferences for venues expressed in check-ins capture cultural differences among users \cite{silvaICWSM14}. Thus, differences of gender preferences for venues expressed in check-ins might also reflect different cultural patterns. In this direction, our methodology might be a useful tool to capture this particular aspect of a certain culture, helping to leverage new types of applications, as discussed in the next section.}

\section{Applications}\label{secApplications}

Many  applications could benefit from our methodology to study gender preferences for venues. Some of them are:

\noindent \textit{Insights for policy-makers}: Policy-makers could use the knowledge about gender preferences for venues to  identify existing problems, and obtain insight into possible solutions for them, such as  effective policies  for gender differences reduction in certain regions or venues of the city.

\noindent\textit{New recommendation systems: } The knowledge about cultural gender preferences for venues in a given city, neighborhood, or category of venues could be exploited in the design of new  location recommendation services that take into account these preferences. These  services could help tourists and residents find places of interest (e.g., where to \willi{go out} in an unknown environment).

\noindent\textit{Understanding Consumers: } Business owners and marketers could use the valuable insights about cultural gender preferences of specific venues or categories of venues, to promote more efficient advertisement.

Next, we present more details of an application that demonstrate one possibility to explore  gender preferences for venues.

\subsection{Areas with similar gender popularity} \label{secclustering}

We here illustrate one particular application that aims at identifying groups of similar urban areas according to the  degree of gender difference observed in the preference for different (types of) places located in those areas, where  gender difference is inferred from the cross-gender popularity differences. As argued above, such popularity differences \thiago{might} reflect different cultural patterns. Thus, by   clustering  regions based on them, we aim at identifying groups of regions that share similar cultural traits related to gender preference for venues. This effort is similar to a recent investigation on using check-ins to identify cultural boundaries based on eating and drinking patterns \cite{silvaICWSM14}, although we here explore a different cultural dimension.

Our goal in this section is to further investigate the extent to which our cross-gender popularity differences provide useful information about gender preference for venues in a given region of the real world.  For that, the application we envision  works as follows. \thiago{We estimate the variability $w$ of the cross-gender popularity differences  measured for all venues (in all subcategories) located in the region under study. A large $w$ across the venues  is taken as a sign of large variability in the cross-gender popularity differences\footnote{We note that the cross-gender popularity differences  might be equally large in all venues, resulting in low variability. Our strategy does not catch those cases. However, this pattern  is unlikely to happen in practice, and indeed we did not observe it in our dataset.}.}

\begin{table}[ttt!]
\caption{Clustering of countries.}
\begin{center}
\scriptsize
    \begin{tabular}{  c | p{2.7cm}  || c | p{2.7cm} }
\multicolumn{2}{c||}{$k$=4}  & \multicolumn{2}{c}{$k$=10} \\ \hline \hline
    Cluster & Countries  &  Cluster & Countries \\ \hline
   1 & Saudi Arabia, United Arab Emirates, Kuwait  & 1& Saudi Arabia, Kuwait\\ \cline{3-4}
          & & 2& United Arab Emirates \\ \hline
          2 & Brazil, Mexico, United States, Japan, Malaysia, Thailand, Turkey& 3& Turkey\\ \cline{3-4}
          & &4& Brazil, Mexico\\ \hline
          3 & France, South Korea, United Kingdom&  5&  South Korea   \\ \cline{3-4}
          &&                                             6& Malaysia,Thailand\\ \hline
           4 & Germany, Spain&  7&   Germany, Spain   \\ \cline{3-4}
          &&                                             8&  France   \\ \cline{3-4}
          &&                                             9&   United Kingdom \\ \cline{3-4}
          &&                                             10&   Japan, United States   \\
    \end{tabular}
\end{center}\label{tabcountriesClusterk4}
\vspace{-5mm}
\end{table}

\thiago{To estimate $w$ we consider the Gini coefficient ($g$), which was proposed to describe the income inequality in a population, but it can be used in the study of inequalities in several domains~\cite{gini2012}. A Gini coefficient of zero expresses perfect equality, where all popularity differences values are the same. A Gini coefficient of one expresses maximal inequality among popularity differences values.

Mathematically, $g$ is defined based on the Lorenz curve, which plots, in our context, the proportion of popularity differences ($y$ axis) that is cumulatively expressed by the $x\%$ of subcategories with smaller popularity differences, as shown by Figure \ref{figLorezCurve}. The line at 45 degrees thus represents perfect equality of popularity differences. The Gini coefficient can then be thought of as the ratio of the area that lies between the line of equality and the Lorenz curve over the total area under the line of equality. Based on Figure \ref{figLorezCurve},  $g = A / (A + B)$.

  \begin{figure}[!hbt]
   \centering
   \includegraphics[width=.40\textwidth]{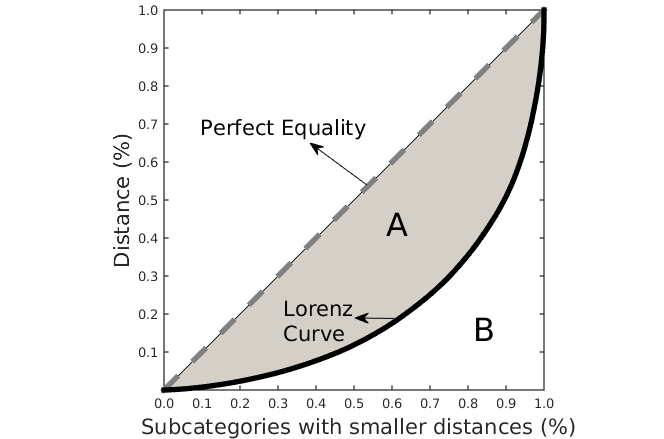}
    \caption{Graphical representation of the Gini coefficient.}
    \label{figLorezCurve}
 \end{figure}

To compute $g$ from an empirical Lorenz curve, one generated by discrete data points (our case), we can use the formula:

\begin{equation}
g= \frac{n+1}{n} - \frac{2\sum_{1}^{n}(n+1-i)x_i}{n\sum_{1}^{n}x_i},
\end{equation}

where the $x_i$ are the popularity differences ordered from least to greatest and $n$ is the number of popularity differences calculated. More details of the Gini Coefficient can be found in~\cite{gini2012}.

}

\begin{table}[ttt!]
\caption{Clustering of cities.}
\begin{center}
\scriptsize
    \begin{tabular}{  c | p{2.7cm}  || c | p{2.7cm} }
\multicolumn{2}{c||}{$k$=\willi{10}}  & \multicolumn{2}{c}{$k$=2} \\ \hline \hline
    Cluster & Cities  &  Cluster & Cities \\ \hline
1&New York, Chicago   &  1 & New York, Chicago, San Francisco, Paris,  Sao Paulo, \\ \cline{1-2}
2&Sao Paulo, Rio de Janeiro, Belo Horizonte & & Rio de Janeiro, Belo Horizonte, Tokyo, Osaka, London, Mexico City\\ \cline{1-2}
3& Johor Bahru, Riyadh, Jeddah & & \\ \cline{1-2}
4& Tokyo, Osaka  & & \\ \hline
5&Kuala Lumpur, Bangkok & 2 & Kuala Lumpur, Johor Bahru, Istanbul, Ankara,\\ \cline{1-2}
6& Istanbul, San Francisco & &  Izmir, Riyadh, Jeddah, Bangkok\\ \cline{1-2}
7&Ankara, Izmir & & \\ \cline{1-2}
8&London& & \\ \cline{1-2}
9&Mexico City&& \\ \cline{1-2}
10&Paris && \\
    \end{tabular}
\end{center}\label{tabcitiesClusterk8}
\vspace{-5mm}
\end{table}

Given a set of regions $R$, we use the Gini metric to  estimate the variability of the  cross-gender popularity differences for individual venues of each subcategory analyzed in each region $r \in R$.  We then represent each region $r$ by a cultural gender preference vector, $G_r=\{g^{S_1},g^{S_2},...,g^{S_n}\}$, where $g^{S_i}$ is the Gini coefficient computed for subcategory $S_i$, and $n$ is the total number of subcategories analyzed in all regions \novo{($n$=299, all subcategories considered.)}. We assume $g^{S_i}$=0 if subcategory $S_i$ was not analyzed in region $r$ due to the lack of enough data. Finally, we use the  $k$-means algorithm (with cosine distance) to cluster the regions in the space defined by $G_r$.

We tested this idea by clustering the 15 countries analyzed. First, we used   $k=4$, as the countries are located in $4$  distinct geographic regions of the world. Table~\ref{tabcountriesClusterk4} shows the identified  clusters. Some groupings are expected according to common sense. For example, all the Arab countries were grouped together, possibly because they share many cultural similarities regarding  female habits. Yet, the table also reveals possibly unexpected results, such as the greater similarity of South Korea with European countries.  Similarly, Thailand, Malaysia, and Turkey are grouped together with Brazil, Mexico, Japan, and United States. Despite the geographic (and perhaps also cultural), distance between some of the countries, they share   similar patterns in cross-gender popularity differences, which might be a reflection of similar social conditions. \thiago{In order to further investigate these results, we identified $k=10$ clusters, results also shown in Table~\ref{tabcountriesClusterk4}.  In this new grouping, UK, France, South Korea  and Turkey represent a cluster by themselves, and Thailand and Malaysia is now a cluster, leaving Brazil and Mexico as another cluster. This result reinforces the suggestion that our data might indeed represent characteristics of the cultural behavior of the inhabitants of those places.}

\thiago{One could think that the result is correlated with the number of data available in the region of study, since some \willi{of the $k=4$} clusters, such as the one containing Germany, Spain, and France, have a small amount of data. However, if this was the case, South Korea and the United Arab Emirates would also be in the same cluster because they also have a small number of data. In order to further investigate this possible problem, we selected 19 popular cities according to the number of check-ins}, representing distinct regions of the world: New York, Chicago, San Francisco (USA), Sao Paulo, Rio de Janeiro, Belo Horizonte (Brazil), Kuala Lumpur, Johor Bahru (Malaysia), Tokyo, Osaka (Japan), Paris (France), London (UK), Istanbul, Ankara, Izmir (Turkey), Riyadh, Jeddah (Saudi Arabia), Mexico City (Mexico), and Bangkok (Thailand).

Table~\ref{tabcitiesClusterk8} (left) shows the  results of clustering these cities using $k$=$10$, the same number of distinct countries where these cities are located. As we can see, most of the cities from the same country were clustered together. One exception, in this sense, was Istanbul grouped with San Francisco. Perhaps, the behavior of users of those cities is in fact more similar to each other than the other cities studied of the same country. Istanbul, due to the penalty mentioned in Section~\ref{subsec:country_analysis}, presented a \willi{distinct} pattern related to soccer places compared to other cities in the same country. The city is also concerned in promoting gender equality and the empowerment of women \cite{turkeyGender2}, and, maybe,  some of the actions in this direction might have an effect, changing the behavior of inhabitants to be more similar to citizens of San Francisco. Besides that, today, Istanbul has the best record in regards to gender equality among 81 Turkish provinces \cite{turkeyGender}. Another exception was Kuala Lumpur grouped with Bangkok instead of Johor Bahru, which was grouped with Riyadh, Jeddah. The fact that Kuala Kuala Lumpur and Bangkok are bigger and more cosmopolitan cities might help to explain this clustering.

Note that by forcing the grouping into only  2 clusters (Table~\ref{tabcitiesClusterk8} - right), our strategy clearly distinguishes cities where most inhabitants have an Islamic tradition (cluster 2), which tends to shape a common cultural gender behavior, from the others. \thiago{Our results suggest that the degree of gender preferences for venues might capture important aspects of gender inequality. \willi{Countries being in the same cluster were classified by sociologists with a similar gender inequality in the Gender Inequality Index (GII)}. We further investigate this question in the next section.  }

\section{Comparison with Official Indices}\label{secValidation}

\pe{Gender inequality can be defined as allowing people different opportunities due to perceived differences based solely on issues of gender \cite{parzialeSage}. This is a broad and complex definition and some initiatives attempt to measure it across different countries, such as the Gender Inequality Index (GII). }  GII is an index for measurement of gender \pe{inequality} developed by the United Nations Development Programme (UNDP), being perhaps the most important study in this area. The index was introduced in the 2010 Human Development Report and we use in this study the 2014 index.  GII is a value ranging from $0$ (no perceivable inequality) to $1$ (extreme inequality) reflecting the inequality between men and women in a given country. It is currently calculated for over $150$ countries, which are ranked by the computed values. More details on calculation of  GII can be found in~\cite{undp2014}.

\pe{We hypothesize that gender preferences for venues expressed in our data might reflect less contact between different genders (recall that we discarded categories that have many subcategories with expected biases towards a particular gender, e.g., Men's Store). This could affect networking opportunities and keep the \say{glass ceilings} in society impermeable, aspects captured by studies of gender inequality such as GII.} \thiago{In this section, we investigate to which extent gender preferences for venues are related to gender \pe{inequality}. \pe{To do that}, we compare the results of our methodology with GII using the cultural gender preference vector, $G_r$, for a country $r$ considered in this study. For that, we rank for a given country $r$ all other countries according to a certain distance towards $r$. In the case of GII values we use euclidean distance and for our vector, we use cosine distance. For example, assuming that $r=Brazil$, we compute the euclidean distance from  GII value for Brazil to all other GII values for the other countries. After that, we compute the cosine distance from the vector representing Brazilians' preferences ($G_{brazil}$) to all other preference vectors for other countries.} Then, we compute a Spearman's rank correlation coefficient $\rho$ \cite{jain2008art} between these two ranks, for each country \pe{(see Appendix \ref{app:a} for more details)}. The idea is to verify if the most similar (and distinct) countries to a particular country in GII, for example, Brazil, are ranked similarly when we use the dimensions computed by our approach.

\pe{Furthermore, in order to verify if the observed relations are more pronounced for gender issues captured by GII, we also make the same comparison explained above using Human Development Index (HDI) and random data, replacing GII in the comparison. HDI is a composite statistic of life expectancy, education, and per capita income indicators. More details about how it is calculated can be found in \cite{undp2014}. In this study, we used HDI from 2014, the same year of our data collection. Since GII includes different dimensions than HDI, it cannot be interpreted as a loss or gain in HDI itself, i.e, it is unrelated to gender. To generate random data we randomly ordered the considered countries. Let $V$ represent a particular rank, in our case we use the values for GII in Table \ref{tabGIIeECI} from Appendix \ref{app:a}, where each line represents a country. We use a function $f$ to perform a random permutation in that vector: $V'=f(V)$, where $V'$ represent a particular permutation of $V$. We created $100$ random ranks: $\mathcal{R}=\{V'_1,V'_2,..V'_n\}$, where $n=100$. We compared every $V'_i \in \mathcal{R}$ with our data, resulting in $100$ $\rho$ correlation values. }



The results are shown in Table \ref{tabGIIcorrelation}. The first column lists  the countries considered, while the second to fifth show the correlation performed $\rho$ and it's respective $p$-$value$, for GII and HDI. We highlight in bold all the coefficients that are \novo{positive} and statistically significant, i.e., with a $p$-$value < 0.05$. \thiago{For example, the first line \pe{for GII} presents the result of the Spearman correlation from the two ranks produced in the example aforementioned for Brazil. In other words, the rank produced of distances from Brazil to the other studied countries for GII values and our preference vectors has a Spearman correlation value of $0.665$, and this value is significant.} \pe{The sixth column represent a 99\% confidence interval of the mean $\rho$ relative to $\mathcal{R}$.}

  \begin{table}[t!]
  \caption{The correlation coefficient $\rho$ (and its p-value) between the rank of similarity generated from GII and HDI with our approach. Significant and positive correlations are rendered in bold.}
  \begin{center}
      \begin{tabular}{  c  c  c | c c | c}
      &\multicolumn{2}{c}{GII} & \multicolumn{2}{c}{HDI}& Random\\
      Country & $\rho$ & p-value &  $\rho$ & p-value & Confidence interval (99\%) of $\rho$ \\ \hline
      Brazil  &      \textbf{0.665} &   \textbf{0.011}    &     \textbf{0.573}  & \textbf{ 0.035}   & (-0.051, 0.071)\\
 France &     \textbf{0.551} &  \textbf{ 0.043}   &     0.520  &  0.059   & (-0.047, 0.103)\\
  Germany &     0.134  &  0.648   &     0.024   & 0.939   & (-0.074, 0.058)\\
  Japan &    -0.569  & 0.036    &    -0.564    & 0.038    & (-0.037, 0.093)\\
  Kuwait  &    \textbf{ 0.709} &   \textbf{0.006}   &     \textbf{0.564}  & \textbf{ 0.038}   & (-0.098, 0.044)\\
  Malaysia  &    -0.345   & 0.227   &     \textbf{0.670}   & \textbf{0.010}   & (-0.070, 0.071)\\
  Mexico  &     \textbf{0.589}  & \textbf{ 0.026}   &     0.446  &  0.111   & (-0.090, 0.049)\\
  Saudi Arabia  &     \textbf{0.558}  &  \textbf{0.037}   &    -0.277 &  0.337    &(-0.152, -0.002)\\
  South Korea   &     \textbf{0.653} &   \textbf{0.011}   &     0.556 &  0.050    & (-0.014, 0.117)\\
  Spain &     \textbf{0.547}  & \textbf{ 0.045}   &     0.363&   0.202    & (-0.067, 0.072)\\
  Thailand  &     \textbf{0.675} &  \textbf{ 0.008}   &     \textbf{0.758}  & \textbf{0.002}    & (-0.081, 0.057)\\
  Turkey  &     \textbf{0.753} &  \textbf{ 0.002}   &     \textbf{0.661 }&  \textbf{0.012}    & (-0.079, 0.043)\\
  UAE &    -0.116  & 0.693    &     0.314   & 0.273   & (-0.111, 0.034)\\
  United Kingdom  &     0.107   & 0.715   &     0.187   & 0.522   & (-0.017, 0.126)\\
  United States &     0.279  &  0.333   &    -0.516   & 0.061   & (-0.108, 0.033)\\
      \end{tabular}
  \end{center}
  \label{tabGIIcorrelation}
  \end{table}

Note in Table \ref{tabGIIcorrelation} that a majority of countries show a positive and significant correlation $\rho$ between our gender preference measure with the GII (9 out of 15 countries). In contrast, fewer countries (5 out of 15) have a positive and significant correlation with the HDI. In addition, most of the positive correlation values are higher for the GII case. Random rankings show no correlation (i.e., $\rho$ close to $0$), as expected. The results suggests the outcomes observed are not explained by a general cultural similarity between countries. Besides, they indicate that cross-gender popularity differences, relying solely on check-in data, \thiago{ might capture} important aspects of gender inequality that emerge in sophisticated studies, such as  GII.  \novo{It is important to mention that there are cases where the proposed method does not seem to be related to the GII. For instance, we can find a significant negative correlation for the case of Japan, fact that also happend in the correlation with HDI. Despite of that,} the results suggest that our proposed methodology \pe{could be exploited} to complement existing methodologies to study gender inequalities, \pe{for instance, as an additional dimension}. \pe{However, further research is needed}.

\section{Limitations}\label{secLimitations}

There are several possible reasons for results observed in the comparison (Section \ref{secValidation}) and also in the clustering results (Section \ref{secclustering}). Some countries in our dataset have a small number of users (and check-ins), possibly reflecting a lower adoption of Foursquare among those countries' inhabitants. This is a limitation of our dataset, which covers only seven days. \thiago{A dataset spanning a longer period would most certainly capture a larger fraction of the population of those countries, although the adoption rate imposes inherent constraints. Besides that, there might be more accurate methods than the Gini coefficient to generate the cultural gender preference vector, other metrics could also be tested aiming to improve the comparison results}. Yet, our methodology also has limitations. Take, for instance, Saudi Arabia, where the same place may have exclusive sectors for men and women, such as restaurants with segregated service and eating zones, and shopping malls with dedicated floors for women (as in the Kingdom Centre\footnote{http://kingdomcentre.com.sa/ladies.html.}). The gender segregation in those places is very high. Yet, our approach is not able to capture the correct level of segregation since those gender-specific sectors and zones are not distinguished as different venues on Foursquare.

\thiago{Besides that, our methodology assumes that the gender information given by users on their profile page are correct. This might not be a significant problem since there is evidence that  users provide correct gender information in their online profiles. Burger et al. \cite{Burger2011} studied user gender on Twitter considering gender information shared by users in external blog accounts associated with their Twitter account. This association enabled an experiment verifying that cues in Twitter profile descriptions, e.g. \say{mother of 3 children},  tend to be consistent with gender information in the blog. This may indicate that people who misrepresent their gender are consistent across different aspects of their online presence. Linked to that, our proposed methodology also does not tackle the case where users do not fit in either male or female gender, as shown by \cite{deLasCasas2014}. Our methodology also does not treat pollution, e.g. fake accounts. In this particular case, techniques to increase data quality could improve the results \cite{Ghosh2012,Gupta2013,Yang2014}. }

\section{Conclusions and Future Work}\label{conclusions}

We have proposed a methodology to identify  gender differences in preferences for specific venues in urban regions by analyzing user check-in data on Foursquare. We illustrated the use of our methodology by applying it to identify statistically significant cross-gender differences in  preferences for  venues, at both country and city levels. Some of these significant differences reflect well-known cultural patterns, while others could be explained by particular aspects of the venues identified after manual research.

We also gathered evidence that our methodology offers useful information about gender preference for venues in a given region in the real world. This result suggests that, despite limitations and biases that might exist in our data, our methodology could be a useful tool to support faster and cheaper large-scale studies on gender preferences for venues.

By exploiting our cross-gender preferences for venue differences,  business owners could gain valuable insights about their customers; venue recommendations could become more culturally-aware, as men and women may have different preferences in regions with distinct cultures; and data-intensive sociological studies about gender preferences for venues could be done in less time, with larger sample sizes, and on regions with arbitrary spatial granularities.

As future work, we intend to validate our methodology with other LBSN datasets and other data about gender preferences for venues collected in a traditional (offline) fashion. \pe{Besides that, we envision to investigate how the proposed methodology could be exploited to complement existing methodologies to study gender inequalities.} We also plan to investigate other applications that can benefit from our results, and expand our methodology to add a temporal dimension, thus capturing temporal variations in cross-gender preferences for venues that might exist.

\section*{Acknowledgments}
This work was partially supported by the FAPEMIG-PRONEX-MASWeb project -- Models, Algorithms and Systems for the Web, process number APQ-01400-14, as well as by the National Institute of Science and Technology for the Web (INWEB), Capes, CNPq, FAPEMIG, and Funda\c{c}\~ao Arauc\'aria.

\section*{Competing interests}

The authors declare that they have no competing interests.

\section*{Authors' contributions}

WM, THS, JMA, AAFL conceived, designed, and coordinated the study; WM and THS carried out data processing; WM and THS performed statistical analysis and visualization of results. All the authors interpreted the results, wrote the manuscript and gave the final approval for publication.

\bibliographystyle{spmpsci}      
\bibliography{referencias}   


\newpage
\appendix
\section{- Details About the Comparison with Official Indices} \label{app:a}

\pe{This appendix shows extra information about the comparison with official indices performed in Section \ref{secValidation}. The data for the Gender Inequality Index and Human Development Index were obtained on the UNDP website (hdr.undp.org). All data refer to the year of 2014. For reference, data for each country studied in this work are presented in Table \ref{tabGIIeECI}.

To perform the comparison considered in Section \ref{secValidation} we have to rank for a given country $r$ all other countries according to a certain distance towards $r$. To illustrate this process, consider $r=Brazil$. The first step is to calculate the euclidean distance vector $D1_r$ from Brazil to all other countries according to GII\footnote{For simplicity we consider in this example only data for GII, but the same procedure has to be performed when considering HDI or random data.}. In other words, we compute the pairwise euclidean distance between pairs of country data. According to our example, Brazil has GII value of $0.457$  (Table \ref{tabGIIeECI}), and we have to compute the distance for all other countries. The result for this example is  $D1_{Brazil} = \{0, 0.369, 0.416, 0.324, 0.070, 0.248, 0.084, 0.173, 0.332, 0.362, 0.077, 0.098, 0.225, \\0.280, 0.177\}$.

After that, we compute the cosine distance\footnote{One minus the cosine of the angle between the considered vectors.} $D2_r$ from the vector representing Brazilians' preferences ($G_{brazil}$) to all other preference vectors for other countries. According to our example $D2_r= \{0,0.754,0.757,0.414,0.556,0.328,0.249,0.563,0.795,0.73,0.324,0.379,0.795,0.601,\\0.378\}$.   Then, we compute a Spearman's rank correlation coefficient $\rho$ \cite{jain2008art} between these two ranks, for each country. But, before that we disregard the distance from $r$ itself, which in our example is located in the first position of the distance vectors. The correlation coefficient $\rho$ for this example, as shown in Table \ref{tabGIIcorrelation}, is $0.66$ (with a p-value of $0.01$).  }

\begin{table}[ht!]
\caption{Considered data for Gender Inequality Index and Human Development Index.}
\begin{center}
    \begin{tabular}{ c  c|  c }
Country &GII value& HDI value\\ \hline
Brazil  & 0.457 & 0.755 \\
France  & 0.088 & 0.888 \\
Germany & 0.041 & 0.916 \\
Japan & 0.133 & 0.891 \\
Kuwait  & 0.387 & 0.816 \\
Malaysia  & 0.209 & 0.779 \\
Mexico  & 0.373 & 0.756 \\
Saudi Arabia  & 0.284 & 0.837 \\
South Korea   & 0.125 & 0.898 \\
Spain & 0.095 & 0.876 \\
Thailand  & 0.38  & 0.726 \\
Turkey  & 0.359 & 0.761 \\
United Arab Emirates  & 0.232 & 0.835 \\
United Kingdom  & 0.177 & 0.907 \\
United States & 0.28  & 0.915 \\

 \end{tabular}
\end{center}
\label{tabGIIeECI}
\end{table}

\end{document}